\documentclass[a4, useAMS,usenatbib]{mn2e}

\pdfoutput=1
\usepackage[pdftex]{graphicx}
\usepackage[pdftex]{color}
\usepackage[colorlinks,bookmarks]{hyperref}
\definecolor{linkblue}{rgb}{0,0,0.8}
\definecolor{linkgreen}{rgb}{0,0.5,0}
\hypersetup{linkcolor=linkblue, citecolor=linkgreen, urlcolor=linkblue}
\usepackage[usenames,dvipsnames]{xcolor}
\usepackage[T1]{fontenc}
\usepackage[utf8]{inputenc}
\usepackage{lmodern}
\usepackage[english]{babel}
\usepackage{bm}
\usepackage{verbatim}
\usepackage[us]{datetime}
\usepackage{amsmath, amssymb}

\def\rd{{\rm d}}

\newcommand{\real}{\mathcal D}
\newcommand{\mock}{\mathcal M}

\newcommand{\post}{\mathfrak L}
\newcommand{\prio}{\mathcal P}
\newcommand{\like}{\mathcal L}
\newcommand{\likem}{\mathcal L_{\rm max}}
\newcommand{\evi}{\mathcal E}

\newcommand{\brat}{\mathcal B}
\newcommand{\tprio}{\tilde \theta}
\newcommand{\tpriobm}{\boldsymbol {\tilde \theta}}
\newcommand{\thetahatbm}{\boldsymbol {\hat \theta}}

\newcommand{\omegam}{\Omega_{m0}}
\newcommand{\sigmaextra}{\sigma_{\rm int}^{\rm extra}}
\newcommand{\omegal}{\Omega_{\Lambda0}}

\newcommand{\thetabm}{\boldsymbol \theta}
\newcommand{\Sigmabm}{\boldsymbol \Sigma}

\newcommand{\mbm}{\boldsymbol m}
\newcommand{\xbm}{\boldsymbol x}
\newcommand{\vbm}{\boldsymbol v}
\newcommand{\Abm}{\boldsymbol A}
\newcommand{\Fbm}{\boldsymbol F}
\newcommand{\Pbm}{\boldsymbol P}
\newcommand{\Lbm}{\boldsymbol L}

\title[Internal Robustness]{Internal Robustness: systematic search for systematic bias in SN Ia data}

\author[L.~Amendola, V.~Marra and M.~Quartin]{Luca Amendola$^{1}$, Valerio Marra$^{1}$ and Miguel Quartin$^{2}$\\
$^{1}$Institut für Theoretische Physik, Universität Heidelberg, Philosophenweg 16, 69120 Heidelberg, Germany\\
$^{2}$Instituto de Física, Universidade Federal do Rio de Janeiro, CEP 21941-972, Rio de Janeiro, RJ, Brazil}

\begin{document}
%%%%%%%%%%%%%%%%%%%%%%%%%%%%%%%%%%%%
%%%%%%%%%%%%%%%%%%%%%%%%%%%%%%%%%%%%

\date{Accepted XXX. Received XXX; in original form XXX}

\pagerange{\pageref{firstpage}--\pageref{lastpage}} \pubyear{2012}

\maketitle

\label{firstpage}

\begin{abstract}
A great deal of effort is currently being devoted to understanding,
estimating and removing systematic errors in cosmological data. In the
particular case of type Ia supernovae, systematics are starting to dominate the
error budget. Here we propose a Bayesian tool for carrying out a systematic
search for systematic contamination. This serves as an extension to the standard
goodness-of-fit tests and allows not only to cross-check raw or processed data
for the presence of systematics but also to pin-point the data that are most likely
contaminated. We successfully test our tool with mock catalogues and conclude
that the Union2.1 data do not possess a significant amount of systematics.
Finally, we show that if one includes in Union2.1 the supernovae that originally
failed the quality cuts, our tool signals the presence of systematics at over 3.8-$\sigma$ confidence level.
\end{abstract}


%%%%%%%%%%%%%%%%%%%%%%%%%%%%%%%%%%%%
%%%%%%%%%%%%%%%%%%%%%%%%%%%%%%%%%%%%
\section{Introduction} \label{intro}
%%%%%%%%%%%%%%%%%%%%%%%%%%%%%%%%%%%%
%%%%%%%%%%%%%%%%%%%%%%%%%%%%%%%%%%%%

The best evidence for the accelerated expansion of the universe still comes,
after 15 years from the earliest results~\citep{Riess:1998cb,Perlmutter:1998np},
from the supernovae Ia (SNIa). There are now several hundreds SNIa useful for
cosmological purposes, ranging in distance up to $z\approx1.7$. The SNIa have
been compiled in different
datasets~\citep{Kowalski:2008ez,Hicken:2009dk,Lampeitl:2009jq,Guy:2010bc,
Conley:2011ku,Suzuki:2011hu}, taking into account different sets of possible
systematics and making use of two different approaches to standardize these
primordial candles~\citep{Jha:2006fm,Guy:2007dv}. Nevertheless, every analysis
performed on these datasets confirms that a present cosmic acceleration explains
satisfactorily the data. The same conclusion is now supported also by several
other lines of evidence, such as measurements of the Baryonic Acoustic
Oscillations (BAO)~\citep{Eisenstein:2005su,Blake:2011wn}, of the anisotropies of
the Cosmic Background Radiation (CMB)~\citep{Komatsu:2010fb} and of the age of
the oldest stars known~\citep{Jimenez:1996at,Carretta:1999ii,Hansen:2002ij}.

The recent increase in the number of observed supernovae is also driving a huge
effort to understand and control possible sources of systematics that may
undermine the progress in the cosmological interpretation. A recent
analysis~\citep{Suzuki:2011hu} claims indeed the systematic uncertainties are
already larger than the statistical ones and the issue will be much more
important in the near future as we forecast an increase in the number of
observed supernovae by 1 order of magnitude in the next $\sim5$ years (for
example with the Dark Energy Survey, see~\citealt{Bernstein:2009ue}) and by 2 orders of
magnitude in the next $\sim15$ (for example with the Large Synaptic Survey
Telescope, see~\citealt{Abell:2009aa}). It is thus necessary to continue investigating
the SNIa datasets in search of such systematic effects and of additional
cosmological information. On one side, in fact, we are already aware of many
effects that could come into play to alter the SNIa apparent magnitude:
contamination from non-Ia supernovae, dust absorption in both host galaxy and
Milky Way, gravitational lensing distortions, local velocity flows \emph{et
cetera}; not to count systematics which arise from selection effects
\citep{Kainulainen:2009dw, Kainulainen:2010at, Kainulainen:2011zx} and from
relying solely on photometry (typically in just a few bandpasses) and the flux
reference of such filters (for a review, see for instance~\citealt{Howell:2010vd}).
On the other side, non standard cosmological models might affect our parameter
estimation: for instance, any anisotropy in the expansion rate
would show up as an anisotropy in the SNIa cosmological parameters. There have
been of course many searches for such systematic biases. All of them, however,
assume a specific effect (say, gravitational lensing as in~\citealt{Amendola:2010ub} or
cosmological anisotropy as in~\citealt{Koivisto:2010dr,Antoniou:2010gw,Colin:2010ds}) and
test whether this effect is enough to make some SNIa incompatible with the
others. In other words, one proceeds by testing a specific prejudice.

In this paper we propose an alternative approach. We wish to perform a
systematic search of biases without having any preferred selection criteria. In
other words, we try to answer the following question: is there any subset of
SNIa that is statistically incompatible with the others? That is, is there a
subset of SNIa that could be described by parameters which are incompatible with
those that describe the other SNIa? In a sense, this is a direct generalization
of the search for outliers. Instead of searching for single outliers, i.e.~SNIa
that appear statistically incompatible with the others (say, some parameters
that describe their light curves are too far off from the others or their
distance moduli just end up very far from the overall Hubble diagram), we search
for subsets of, say, dozens of SNIa at once whose parameters are incompatible
with the others. In other words, we search for heteroscedasticity in the SNIa data.
As will be shown in Section~\ref{dirs}, the proposed
generalization reduces to the standard outlier search in the limit in which the
whole data is divided into two complementary datasets, one of which contains a
single element.

The standard tool to compare whether a particular dataset is compatible with a
proposed model is the~\emph{goodness-of-fit} test, which gives well defined
probability statements about such agreement. However, the information obtained
is limited, and this simple analysis may hide problems in both data and model.
For instance, if a given model parameter affects only a small fraction of the
data, even if such parameter turns out to disagree with this fraction the
goodness-of-fit may still claim an overall good fit. To address this issue, a
modification of the test, dubbed \emph{parameter goodness-of-fit}, was proposed
in~\citealt{Maltoni:2002xd} and later extended in~\citealt{Maltoni:2003cu}. The
parameter goodness-of-fit method nevertheless still relies on comparisons of
$\chi^{2}$ values which are only sensitive to the local minimum, and not to the
entire likelihood. Here, instead, we will adopt a fully Bayesian approach so as
to use all the information available, e.g.~a possible overlapping of the
likelihoods surfaces.
We dub \emph{internal robustness} the fundamental quantity evaluated in this
method. The name is motivated by the analogous quantity
\emph{robustness} which was recently defined by \citealt{March:2011rv} (and originally introduced in \citealt{Marshall:2004zd}) in a context in which the two
datasets refer to different observational probes (and which
we will henceforth refer to as \emph{external robustness}, for differentiation).
 In particular, \citealt{March:2011rv} showed that the robustness is an
estimator ``orthogonal'' to the Figure of Merit, which is sensitive to the
relative orientation of the two probes but not to the distance between the
two-experiment confidence regions.
%As it will be shown in Section~\ref{spir}, the internal robustness can be formally approximated by the parameter goodness-of-fit.

This paper is organized as follows. In Section~\ref{sec:robust} we will
introduce the formalism of the internal robustness.
In Section~\ref{sec:scanna} we will describe how we will systematically search
for bias in SNe data.
In Section~\ref{results} we will show the results relative to a biased test
catalogue, the Union2.1 catalogue augmented with the supernovae that did not
pass the quality cuts, and the actual Union2.1 dataset. As we will see, our
method will be able to detect the systematic bias in the first two catalogues.
Moreover, our analysis does not show signs of systematic effects in the actual
Union2.1 catalogue.
Finally, we will give our conclusions in Section~\ref{sec:conco}, and explain
some technical details in Appendix~\ref{app:details}.

We will adopt the following notation.
Bold face will distinguish a vector $\bm x$ or matrix $\boldsymbol{A}$ from their components $x_{i}$ and $A_{ij}$, a superscript $t$ will denote a transposed vector or matrix, $|\boldsymbol{A}|$ will represent the determinant of a matrix $\boldsymbol{A}$, and a hat will indicate the best-fit value of the corresponding quantity.

\section{Formalism}

\label{sec:robust}

\subsection{Bayesian evidence and its Fisher approximation}

Let us first of all recall some statistical definitions in the Bayesian
context~\citep{Trotta:2008qt,2010deto.book.....A}. The Bayesian \emph{evidence} is defined as
\begin{equation}
    \evi({\bm{x}};M)=\int \like ({\bm{x}};\thetabm^{M}) \prio(\thetabm^{M})\,\rd^{n}\thetabm^{M}\,,\label{eq:likbaym}
\end{equation}
where ${\bm{x}}=(x_{1},x_{2},...,x_{N})$ are $N$ random data, $\thetabm^{M} = (\theta_{1},\theta_{2},...,\theta_{n})$ are $n$ theoretical parameters that describe the model $M$, $\like$ is the likelihood function, and $\prio$ is the prior probability of the parameters $\thetabm^{M}$. If $\prio(M)$ is the prior on a particular model $M$, we can use Bayes' theorem %again
to write
\begin{equation}
    \post(M;{\bm{x}})=\evi({\bm{x}};M)\frac{\prio(M)}{\prio({\bm{x}})}\,,
\end{equation}
i.e.~the posterior probability $\post$ of having model $M$ given the data. We can finally use the latter equation to compare quantitatively two models taking the ratio of their probabilities (so that $\prio({\bm{x}})$ cancels out):
\begin{equation}
    \frac{\post(M_{1};{\bm{x}})}{\post(M_{2};{\bm{x}})}=\brat_{12}\frac{\prio(M_{1})}{\prio(M_{2})}\,,
\end{equation}
where we introduced the Bayes ratio (sometimes referred to as Bayes factor)
\begin{equation}
    \brat_{12}
    %=\frac{\int \like({\bm{x}};\thetabm^{M_{1}})\prio(\thetabm^{M_{1}})\rd^{n_{1}}\thetabm^{M_{1}}}{\int \like({\bm{x}}_{};\thetabm^{M_{2}})\prio(\thetabm^{M_{2}})\rd^{n_{2}}\thetabm^{M_{2}}}
    = {\evi({\bm{x}};M_{1}) \over \evi({\bm{x}};M_{2})}
    \,.\label{eq:bayesratio}
\end{equation}
Often, however, one assumes that $\prio(M_{1})=\prio(M_{2})$ and we adopt this choice here. A Bayes ratio $\brat_{12}>1$ ($<1$) says that current data favors the model $M_{1}$ ($M_{2}$). As we will see in the next Section the Bayes ratio will be central in the definition of \emph{internal robustness}.

Suppose now the likelihood is gaussian in the data with covariance matrix $\boldsymbol{\Sigma}$ and expected means $m_{i}$. Then
\begin{equation} \label{like1}
    \like\,=\,(2\pi)^{-N/2}\,|\boldsymbol{\Sigma}|^{-1/2}\, e^{-\frac{1}{2} \chi^{2}}\,,
\end{equation}
where the $\chi^{2}$ is defined as:
\begin{equation}
    \chi^{2} \equiv (x_{i}-m_{i})^{t} \, \Sigma_{ij}^{-1}(x_{j}-m_{j}) \,.
\end{equation}
The best-fit (minimum) $\chi^{2}$ is then
\begin{equation}
    \hat \chi^{2}\,=\,(x_{i}-\hat{m}_{i})^{t} \, \Sigma_{ij}^{-1}(x_{j}-\hat{m}_{j}) \,,
\end{equation}
where $\hat{m}_{i}$ are the best-fit means. The maximum of the likelihood is then:
\begin{equation}
    \likem \,=\,(2\pi)^{-N/2}|\boldsymbol{\Sigma}|^{-1/2}e^{-\frac{1}{2}\hat \chi^{2}}\,,\label{eq:lmax}
\end{equation}
so that we can rewrite Eq.~(\ref{like1}) as
\begin{equation}
    \like\,=\, \likem \, e^{-\frac{1}{2}( \chi^{2}- \hat \chi^{2})}\,.
\end{equation}
According to \eqref{eq:likbaym}, the $N$ means $m_{i}$ depend on $n$ parameters $\theta_{k}$, % $k\in\{1,...,n\}$
i.e.~$m_{i}=m_{i}(\thetabm)$.\footnote{Note that to simplify the notation we will drop in the following equations the superscript $M$ of the model in question.} The best-fit values $\hat{m}_{i}$ are then functions of the best fit estimators $\hat \theta_{k}$, i.e.~$\hat{m}_{i}=m_{i}\big(\thetahatbm \,\big)$. Here for simplicity we assume that $\Sigmabm$ does not depend on the parameters, but this assumption can be easily lifted.

Let us assume now that the likelihood can be approximated near $\thetahatbm$ by a Gaussian distributions also \emph{in the parameters}, i.e.
\begin{equation}
    \like({\bm{x}};\thetabm)\,\simeq \,f({\bm{x}};\thetabm)\,\equiv\, \likem \, e^{-\frac{1}{2}(\theta_{i}-\hat \theta_{i} )^{t} \, L_{ij}(\theta_{j}-\hat \theta_{j})}\,,\label{eq:gaussianlhood}
\end{equation}
where $L_{ij}$ in the exponential factor is the inverse of the covariance matrix of the likelihood (or Fisher matrix, see e.g.~\citealt{Bassett:2009uv,2010deto.book.....A}) and where now the data are inside the best-fit estimators $\hat \theta_{i}$. Similarly, we assume a gaussian prior so that
\begin{equation}
    \prio(\theta_{k})=\frac{|\bm{P}|^{1/2}}{(2\pi)^{n/2}}\, e^{-\frac{1}{2}(\theta_{i}- \tprio_{i})^{t} \, P_{ij}(\theta_{j}-\tprio_{j})}\,,
\end{equation}
where $\tprio_{i}$ are the prior means and $\bm{P}$ is the prior matrix. It is now possible to evaluate the evidence analytically. Using the relation
\begin{equation}
    \int\rd^{n}\xbm\; e^{- {1 \over 2} \xbm^{t}\Abm\xbm+\vbm^{t}\xbm}\,=\,  {(2 \pi)^{n/2} \over |\Abm|^{1/2}} \; e^{\frac{1}{2} \vbm^{t}\Abm^{-1}\vbm}\label{eq:gaussianintegral}
\end{equation}
in
\begin{align}
    \evi \;=\; & \int f({\bm{x}};\thetabm)\, \prio(\thetabm)\,\rd^{n}\thetabm\nonumber \\
    \;=\; & \; \likem \frac{|\bm{P}|^{1/2}}{(2\pi)^{n/2}}\int\exp\left[-\frac{1}{2}(\theta_{i}-\hat \theta_{i} )^{t} \, L_{ij}(\theta_{j}-\hat \theta_{j}) \right.\nonumber \\
     & -\left. \frac{1}{2}(\theta_{i}- \tprio_{i})^{t} \, P_{ij}(\theta_{j}-\tprio_{j}) \right]\rd^{n}\thetabm\,, \label{eq:evidence-full}
\end{align}
one finds
\begin{align}
    \evi \,= & \; \likem \,\frac{|{\bm{P}}|^{1/2}}{|{\bm{F}}|^{1/2}}\;\, e^{-\frac{1}{2}\big(\thetahatbm-\tpriobm \big)^{t}\Lbm\Fbm^{-1}\Pbm\big(\thetahatbm-\tpriobm \big)}\,,\label{eq:evidence-v2}
\end{align}
where ${\bm{F}}={\bm{P}}+{\bm{L}}\,$. It will be convenient to rewrite the above equation as
\begin{align}
    \evi\,=\, \likem \,\frac{|{\bm{P}}|^{1/2}}{|{\bm{F}}|^{1/2}} \;\, e^{-\frac{1}{2}\Big(\thetahatbm^{t}\Lbm \thetahatbm\;+\;\tpriobm^{t}\Pbm \tpriobm\;-\;\thetabm'^{t}\Fbm\thetabm' \Big)}\,,\label{eq:evidence}
\end{align}
where $\thetabm'=\Fbm^{-1}\big(\Lbm \thetahatbm+\Pbm \tpriobm \,\big)$. In the limit in which the prior is very broad (so that $P_{ij}\ll L_{ij}$ and therefore $F_{ij}\rightarrow L_{ij}$) the argument of the exponential vanishes and we have simply %\textbf{[MQ: I think the second assumption is irrelevant, as per~\eqref{eq:evidence-v2} it only matters that the prior is broad]}
\begin{equation}
    \evi\,\simeq\,(2\pi)^{-N/2}e^{-\frac{1}{2} \hat \chi^{2}} \frac{|{\bm{P}}|^{1/2}}{|{\bm{L}}|^{1/2}|\mathbf{\Sigma}|^{1/2}}\,.\label{eq:evidence-2}
\end{equation}

\subsection{Definition of internal robustness} \label{dirs}

Eq.~(\ref{eq:likbaym}) and its Fisher approximation Eq.~(\ref{eq:evidence}) allow to compute the evidence $\evi_{\rm tot}$ for a given cosmological model $M_{C}$ and dataset $d_{\rm tot}$. Suppose now that the data are actually coming from two completely different, and therefore independent, distributions. In other words, assume that a subset $d_{2}$ of $d_{\rm tot}$ actually depends on a totally different set of parameters (say, the properties of the SNIa progenitors or galaxy environment), i.e., it is fully described by the systematic parameters of model $M_{S}$. Contrarily, the complementary set $d_{1} = d_{\rm tot} - d_{2}$ is still described by the cosmological parameters of model $M_{C}$. In this case the total evidence can  be written as the product of the individual evidences:
\begin{equation}
    \evi_{\rm ind}=\evi_1 \; \evi_2 \,.
\end{equation}
We can now use the Bayes ratio of Eq.~(\ref{eq:bayesratio}) to quantify which hypothesis is favored. We thus compute\footnote{It is straightforward to generalize Eq.~(\ref{hr1}) to more than two partitions such that $\evi_{\rm ind}=\evi_1 \; \evi_2 \; \evi_3...$}
\begin{equation} \label{hr1}
    \brat_{\rm tot, ind} \;=\;\frac{\evi_{\rm tot}}{\evi_{\rm ind}}
    %=\frac{\evi_{\rm tot}}{\evi_1 \; \evi_2}
    = \frac{\evi(\bm{x}; M_{C})}{\evi(\bm{x}_{1}; M_{C}) \; \evi(\bm{x}_{2}; M_{S})} \,,
\end{equation}
and define
\begin{equation}
    R\;\equiv\;\log \brat_{\rm tot, ind}
\end{equation}
as the \emph{internal robustness}. As discussed in the Introduction this quantity is related to the~\emph{external robustness} originally defined simply as \emph{robustness}  in~\citealt{March:2011rv}.
The model $M_{S}$ may be (and we will do so in the present paper) identified with the cosmological model $M_{C}$, see Section \ref{sec:syst-param} for more details.

The previous equations give the general definition of internal robustness. It is however useful to evaluate analytically $R$ in the Fisher approximation.
Using Eq.~(\ref{eq:evidence}) one finds:
\begin{equation}
\begin{aligned}
    & \brat_{\rm tot, ind} \;=\;\left( \frac{|\Lbm_{1}+\Pbm_{C}||\Lbm_{2}+\Pbm_{S}|}{|\Lbm_{{\rm tot}}+\Pbm_{C}||\Pbm_{S}|}\,
    %\frac{|\Sigmabm_{1}||\Sigmabm_{2}|}{|\Sigmabm_{{\rm tot}}|}
    \right)^{1/2}
    \frac{\likem^{\rm tot}}{\likem^{[1]} \; \likem^{[2]} } \\
    &\!\!\! \times e^{-\frac{1}{2}\Big[
    \thetahatbm^{t}_{\rm tot}\Lbm_{\rm tot} \thetahatbm_{\rm tot}-\thetabm'^{t}_{\rm tot} \Fbm_{\rm tot} \thetabm'_{\rm tot}
    -\sum_{i}\! \big( \thetahatbm^{t}_i \Lbm_i \thetahatbm_i-\thetabm'^{t}_i \Fbm_i \thetabm'_i\big) -\tpriobm_{2}^{t} \Pbm_{S} \tpriobm_{2}
    \Big]} \\
    &\!\!\! \simeq\,\left(\frac{|\Lbm_{1}||\Lbm_{2}|}{|\Lbm_{{\rm tot}}||\Pbm_{S}|}\right)^{1/2}\,\left(\frac{|\Sigmabm_{1}||\Sigmabm_{2}|}{|\Sigmabm_{{\rm tot}}|}\right)^{1/2}\, e^{-\frac{1}{2}\left(\hat \chi_{{\rm tot}}^{2}- \hat \chi_{1}^{2}-\hat \chi_{2}^{2}\right)},
\end{aligned}
\end{equation}
where in the bottom line we simplified using Eq.~(\ref{eq:evidence-2}), i.e., assuming the prior to be much
broader than the likelihoods.
%The subscripts ``$C$'' (``$S$'') refers to the quantities evaluated for the model with cosmological (systematic) parameters.
Notice that the $(2\pi)^{-N/2}$ factors cancel out since $N_{{\rm tot}}=N_{1}+N_{2}$, where $N_{i}$ is the size of the subset $d_{i}$.
So far we have assumed gaussianity of the likelihoods, the existence
of two independent distributions and the use of a very broad prior.
If we make the additional assumption that the data points themselves
are independent of one another (henceforth referred to as ``raw data''),
then the covariance matrix is diagonal, one has $|\Sigmabm_{{\rm tot}}|=|\Sigmabm_{1}||\Sigmabm_{2}|$
and we get the final formula for the internal robustness in the Fisher approximation:
\begin{equation}
R=R_{0}+\frac{1}{2}\log\left(\frac{|\Lbm_{1}||\Lbm_{2}|}{|\Lbm_{{\rm tot}}|}\right)-\frac{1}{2}\left(\hat \chi_{{\rm tot}}^{2}-\hat \chi_{1}^{2}-\hat \chi_{2}^{2}\right), \label{eq:robustness}
\end{equation}
where $R_{0}$ is a constant coming from the unknown determinant of the systematic prior, $R_{0}=-{1 \over 2} \log|\Pbm_{S}|$.

The first factor in Eq.~(\ref{eq:robustness}), formed out of the determinants, expresses Occam's
razor factor of parameter volumes, while the second penalizes $R$
if the two probes are very different from each other (so the hypothesis
that they come from different models, or equivalently that systematics are important, is favored).
We expect, therefore, the internal robustness to be a measure of how much subsets of a dataset overlap: the more they do, the more compatible the two datasets are.

In order to help intuition, we can evaluate $R$ for the simplified
case in which $\,a$) we can neglect the logarithmic part; $\,b$) there
is only one parameter; $\,c$) the errors are all identical ($\sigma_{i}=\sigma$); and $\,d)$
the subset $d_{2}$ consists of a single point $x_{2}$.
%, e.g. a supernova of magnitude $x_{2}$.
Then we have $\hat \chi_{2}^{2}=0$ and for $N_{1} \simeq N_{\rm tot} \gg1$
\begin{equation}
R\approx-\frac{1}{2}\frac{(x_{2}-\hat{m}_{\rm tot})^{2}}{\sigma^{2}}
\end{equation}
i.e.~$R$ reduces to the scatter of $x_{2}$ from the best fit $\hat{m}_{\rm tot}$
evaluated by fitting the remaining $N_1$ elements.
Therefore a large and negative $R$ means $x_{2}$ is an outlier.

\subsection{Statistical properties of internal robustness} \label{spir}

In order to understand the statistical properties of the internal robustness let
us start by fixing the subset $d_{2}$ to some $d_{2}^{*}$ ($d_{1}^{*}$ is just
the complementary subset).
Let us also assume that data come from the cosmological model $M_{C}$ only.
From Eq.~(\ref{eq:robustness}) we can then calculate the probability
distribution function of $R(d_{2}^{*})$ (denoted as eR-PDF).
If one also neglects the logarithmic term (Occam's razor factor), then (in this
very particular case) $R$ becomes the parameter goodness-of-fit test
introduced by \citealt{Maltoni:2002xd}, and it was shown by \citealt{Maltoni:2003cu} that
$R(d_{2}^{*})$ is distributed as a $\chi^{2}$ distribution with $n_{\rm tot}$
degrees of freedom (d.o.f.).
The full eR-PDF is then a modified $\chi^{2}$ distribution with $n_{\rm tot}$
d.o.f., not too distorted as Occam's razor factor is logarithmically suppressed.
If we now drop the assumption that data come from the cosmological model
$M_{C}$, we can use the (fiducial) eR-PDF to assess the significance of a given
value of $R(d_{2}^{*})$, for example of a low value that could indicate that the
dataset is systematics driven.
We remind indeed the reader that in our fully Bayesian context the robustness $R$ is
related to the Bayes ratio of the evidences and a small (large) $R$ disfavors
(favors) the description of the dataset by the cosmological model $M_{C}$ alone.

As far as the internal robustness is concerned, however, we do not intend to fix $d_{2}$ to a particular subset (even if we may still do so if useful). The above way of proceeding is suited indeed to the external robustness, the aim of which is to analyze two different datasets (e.g. CMB and BAO).
The idea behind the internal robustness is instead to keep the total dataset fixed (i.e.~not to be concerned with the statistical distribution of $R(d_{\rm tot})$) and evaluate $R(d_{2})$ for all the possible partitions of $d_{\rm tot}$, thus generating a distribution which we call iR-PDF.
Internal and external robustness have therefore very \emph{different} statistical properties.

The fiducial iR-PDF (assuming that data come from the cosmological model $M_{C}$ only) is a highly nontrivial object. Even if one neglects Occam's razor factor, it is \emph{not} a $\chi^{2}$ distribution with $n_{\rm tot}$ d.o.f., as the sampling of $d_{2}$ is constrained within a fixed realization of $d_{\rm tot}$, on which the iR-PDF depends. As a simple example of the difference between the latter two distributions, the iR-PDF has a compact and discrete support as it is sampled over a finite number of subsets.
%To our knowledge, there is not a simple way to relate $d_{\rm tot}$ to iR-PDF.
We will see in Section \ref{limichi}, however, that for the datasets treated in this paper the binned iR-PDF (neglecting Occam's razor factor) is rather close to a $\chi^{2}$ distribution with $n_{\rm tot}$ degrees of freedom.

The iR-PDF gives the probability that a given value of $R$ is realized among the available subsets.
However, differently from the eR-PDF, the iR-PDF cannot be used to assess the significance of a given value of $R$.
To do so, we need to compute a distribution of iR-PDFs, which we will obtain by evaluating the robustness in many Montecarlo realizations of mock catalogues.
We would like to stress that thanks to this approach our analysis will not be affected by the possible Bayesian penalization of models with many parameters.

%Moreover, the iR-PDF is a distribution of $R$ values sampled over the subsets, which are \emph{not} random variables: the random variable is the realization $d_{\rm tot}$ which is kept fixed. The iR-PDF encapsulates the probability that a given value of $R$ is realized among the available subsets. As a consequence there is not a probability attached to a given value of $R$ that can be used to say if that value is ``normal'' or not. This was not the case of the eR-PDF which truly gives the probability of a random variable to occur. The above reasoning means that we need to compute a distribution of iR-PDFs, in order to assess how likely is the iR-PDF of the given $d_{\rm tot}$. As a distribution of iR-PDFs does not seem to be known, we will address this issue by means of a Monte Carlo analysis, with which the behavior of the object at hand is understood by means of a large controlled input.

%%%%%%%%%%%%%%%%%%%%%%%%%%%%%%%%%%%%
%%%%%%%%%%%%%%%%%%%%%%%%%%%%%%%%%%%%
\subsection{Systematic parameters}

\label{sec:syst-param}

%%%%%%%%%%%%%%%%%%%%%%%%%%%%%%%%%%%%
%%%%%%%%%%%%%%%%%%%%%%%%%%%%%%%%%%%%

Generally speaking, there are two possible choices for the parameters for the subset $d_{2}$. In the first case, which we will adopt in the analysis of this paper, the parametrization is analog to the cosmological one, e.g.~$\{\omegam,\omegal,\alpha \}$, where $\omegam, \omegal$ are the present-day matter and dark-energy parameters and $\alpha$ is combination of the unknown magnitude offset $M_0$ (sum of the SNe absolute magnitudes, of $k$-corrections and other possible systematics) and the Hubble parameter $H_0$: $\,\alpha \equiv M_0 - 5 \log_{10} H_0 10$pc~\citep{2010deto.book.....A}. In this case $\Pbm_{S}=\Pbm_{C}$. This would be the preferred choice if we expect some of the SNe to be better described by different cosmological parameters.
For instance the parameters could differ in different line-of-sight directions (say, 3 different Hubble parameters $H_{x0}$, $H_{y0}$ and $H_{z0}$, as in anisotropic models with shear, see \citealt{Graham:2010hh,Colin:2010ds}) or in different redshift shells (say, $\Omega_{m0}^{\rm in}$ and $\Omega_{m0}^{\rm out}$ as in some inhomogeneous models, see \citealt{GarciaBellido:2008nz,Marra:2011ct,Jackson:2012jt}). We could also erroneously interpret the unknown systematic parameters as the cosmological ones and therefore employ the same parameter names and the same prior function.

In the second case, not to be explored here, one could choose a completely phenomenological parametrization such as
\begin{equation}
    m(z)=\sum_{i=0}^{j}\lambda_{i}f_{i}(z)
\end{equation}
for the theoretical magnitudes. Here the functions $f(z)$ could be arbitrarily chosen, e.g.~$f_{i}(z)=z^{i}$. If $j=2$ we have then $m=\lambda_{0} + \lambda_{1}z + \lambda_{2}z^{2}$. Since the $\lambda$'s are linear parameters, both best fit and Fisher matrix would be analytical. This second choice would be appropriate if one expects some of the SNe  to be %completely
dominated by systematic effects unrelated to cosmology, say because of sample contamination or strong environmental effects.

Restricting ourselves to the first case, we can marginalize over $\alpha$ analytically~\citep{2010deto.book.....A}. Let us define the sums (remember that the covariance matrix $\Sigma$ is diagonal)
\begin{equation}
S_{n}=\sum_{i=1}^{N'}\frac{M_{i}^{n}}{\sigma_{i}^{2}}\,,
\end{equation}
where $N'=N_{1,2}$, $M_{i}\equiv x_{i} -m_i= x_{i} - 5\log_{10} \bar d_{L}(z_{i})$, $\bar d_{L}=  d_{L} H_0$ and $d_{L}$ is the luminosity distance. Marginalizing over the constant offset $\alpha$ we obtain
\begin{equation} \label{lamar}
- \log \like= {1 \over 2} \chi_{\rm mar}^{2} + {1 \over 2} \log\frac{S_{0}}{2\pi} + \sum_{i=1}^{N'} \log\big(\sqrt{2\pi}\sigma_{i}\big) \,,
\end{equation}
where
\begin{equation}
\chi_{\rm mar}^{2}=S_{2}-\frac{S_{1}^{2}}{S_{0}} \,.
\end{equation}
As the likelihood is gaussian in the data, minimizing the marginalized $\chi_{\rm mar}^{2}$ with respect to $\{\omegam,\omegal\}$ gives the same result as minimizing the original $\chi^{2}$ with respect to $\{\omegam,\omegal,\alpha\}$. Therefore, even though we are in a Bayesian context, $\hat \chi_{\rm mar}^{2}$ is still distributed as a $\chi^{2}$ with $N'-3$ degrees of freedom. Note also that if the original, non-marginalized data is independent, then so will be the data marginalized over $\alpha$. To see this, one can rewrite the marginalized likelihood as
\begin{equation}
\like \propto    \exp\left[-\frac{1}{2}\big(\xbm-\mbm \big)^{t}\;\overline{\Sigmabm}^{-1}\;\big(\xbm-\mbm \big)\right],
\end{equation}
 where $\overline{\Sigmabm}^{-1}=\Sigmabm^{-1}-\Sigmabm^{-2}/S_{0}$, which
for a diagonal $\Sigmabm$ is still a diagonal matrix.
%One also has $\big|\overline{\Sigmabm}^{-1}\big|=|\Sigmabm|^{-1}|\mathbf{I}-\Sigmabm^{-1}/S_{0}|$.

Recalling that in general a Fisher matrix is given by
\begin{equation}
L_{pq}=-\frac{\partial^{2}\,\log \like }{\partial\theta_{p}\partial\theta_{q}} \,,
\end{equation}
 we get for the cosmological Fisher matrices $\Lbm_{{\rm tot}}$, $\Lbm_{1}$ and $\Lbm_{2}$
the general expression
\begin{align}
    L_{pq} &\;=\; \frac{1}{2}S_{2,pq}-\frac{1}{S_{0}}\big(S_{1}S_{1,pq}+S_{1,p}S_{1,q}\big)\\
    = & \;\; \sum_{i}\frac{M_{i,pq}}{\sigma_{i}^{2}}\left(M_{i}-\frac{S_{1}}{S_{0}}\right)+\sum_{i }\frac{M_{i,p}M_{i,q}}{\sigma_{i}^{2}} \nonumber\\
    & \;\; -\frac{1}{S_{0}}\sum_{i} \frac{M_{i,p}}{\sigma_{i}^{2}}\sum_{j} \frac{M_{j,q}}{\sigma_{j}^{2}} \nonumber\\
    = & \;\; \frac{5}{\ln10}\sum_{i} \frac{1}{\sigma_{i}^{2}}\left[\frac{\bar d_{Li,p}\bar d_{Li,q}}{\bar d_{Li}^{2}}-\frac{\bar d_{Li,pq}}{\bar d_{Li}}\right]\left(M_{i}-\frac{S_{1}}{S_{0}}\right) \nonumber\\
    +& \,\left[\frac{5}{\ln10}\right]^{2} \! \Bigg[\sum_{i} \frac{\bar d_{Li,p}\bar d_{Li,q}}{\sigma_{i}^{2}\,\bar d_{Li}^{2}}
    -\frac{1}{S_{0}}\sum_{i} \frac{\bar d_{Li,p}}{\sigma_{i}^{2}\bar d_{Li}} \sum_{j} \frac{\bar d_{Lj,q}}{\sigma_{j}^{2}\bar d_{Lj}}\Bigg]. \nonumber
\end{align}

Finally, the robustness in the Fisher approximation is given in this case [see Eq.~(\ref{lamar})] by:
\begin{align} \label{eq:robustnessMa}
    R \,=\,& %-{1 \over 2} \log|\Pbm_{S}|
    \; R_{0}- {1 \over 2} \log(2 \pi) \\
     &+\, \frac{1}{2}\log\left(\frac{S_{0,1}S_{0,2}}{S_{0,{\rm tot}}} \frac{|\Lbm_{1}||\Lbm_{2}|}{|\Lbm_{{\rm tot}}|}\right)-\frac{1}{2}\left(\hat \chi_{{\rm tot}}^{2}-\hat \chi_{1}^{2}-\hat \chi_{2}^{2}\right). \nonumber
\end{align}
Note that in the following results we will express the numerical values of the robustness as $R-R_{0} - \log(2 \pi)$, to which we will refer simply as $R$.

In Section \ref{results} we shall use supernova data provided by the Union2.1~\citep{Suzuki:2011hu} collaboration in the form of a 3-column matrix, each row $\{ z_{i}, x_{i}, \sigma_{i} \}$ consisting of redshift $z_{i}$, distance modulus $x_{i}$ and distance modulus error $\sigma_{i}$. This matrix was computed using the SALT2 method~\citep{Guy:2007dv}, and the nuisance parameters $\alpha$ and $\beta$ controlling stretch and color corrections were fixed to the best-fit values (to wit $\alpha = 0.1219$, $\beta = 2.466$). Used in this way, the supernova data are rigorously not independent, as the values of the distance moduli are obtained after processing the raw data assuming a particular cosmological model (see, e.g.~\citealt{Marriner:2011mf}). As in this first work we mainly aim at presenting the method, we will ignore however such correlations.\footnote{It is worth stressing that the important factor is the independence of the data points, not of their error bars, which even if completely correlated would not affect the results.}
Note, however, that if on one hand this is a caveat for the results that follow, on the other hand it presents an opportunity to cross-check the Union2.1 data (often naively employed in such concise form) for leftover systematics.

\section{Scanning the subsets} \label{sec:scanna}

In order to obtain the full iR-PDF we have to compute $\hat \chi_{1,2}^{2}$ and $\Lbm_{1,2}$ for every partition $d_{1,2}$ of a given dataset of $N_{\rm tot}$ elements. There are
\begin{equation}
    2^{N_{\rm tot}-1}-1\approx10^{0.3\, N_{\rm tot}} %\approx10^{150}
\end{equation}
possible partitions and we find that in the present application a complete scan of all subsets is unfeasible for $N_{\rm tot}\gtrsim20$. The issue then arises of which subset $\Xi$ among all possible partitions to form. We extract from the entire Union2.1 catalogue a number $T$ of subsets $d_{2}$ composed by a number $N_{2}$ of SNe between $N_{2,{\rm min}}$ and $N_{2,{\rm max}}$ chosen at random among all the possible combinations. However, a pure random sampling (i.e.~uniform in the space of all possible partitions) would pick with extremely high probability only the most populated subsets (in our case with $N_{2} \sim N_{2,{\rm max}}$). Since we would like to explore also the smaller subsets, we adjust the selection so as to obtain a distribution approximately uniform in $N_{2}$ (i.e.~approximately equal number of subsets for every value of $N_{2}$). We call this particular set $\Xi(T)$, and the following analysis depends on it. In particular $T$ gives the statistics of the analysis, while the definition of $\Xi$ determines the way the sets have been chosen. We will consider different strategies in forthcoming work.

The upper limit we use is $N_{2,{\rm max}}=N_{\rm tot}/2$. This is due to the fact that, as we are using
the cosmological parametrization, the robustness is symmetric in
the datasets $d_{1},d_{2}$ so that scanning half of the catalogue is
enough.
In order to discuss $N_{2,{\rm min}}$ it is important to stress
that we consider a much larger parameter space than the usual physical
one, as $\omegam$ and $\omegal$ also parametrize the
(possibly cosmology unrelated) systematic parameters. The range we
adopt is $-10<\omegam<10$ and $-20<\omegal<10$, and
we exclude the $\{\omegam,\omegal \}$ region of the parameter
space for which the expansion rate $H(z)$ is negative for $z<2$,
which well accommodates the redshift range of the Union2.1 dataset.
This is a relaxation of the usual no-big-bang excluded region, see Appendix \ref{app:details} for more details.
The value of $N_{2,{\rm min}}$
is then found by demanding that the likelihood of the smaller
subset $d_{2}$ has support within the parameter space considered.
We have found empirically that a value of $N_{2,{\rm min}}=10$ satisfies
on average this requirement.

%%%%%%%%%%%%%%%%%%%%
%%%%%%%%%%%%%%%%%%%%
\begin{figure}
\begin{center}
\includegraphics[width=\columnwidth]{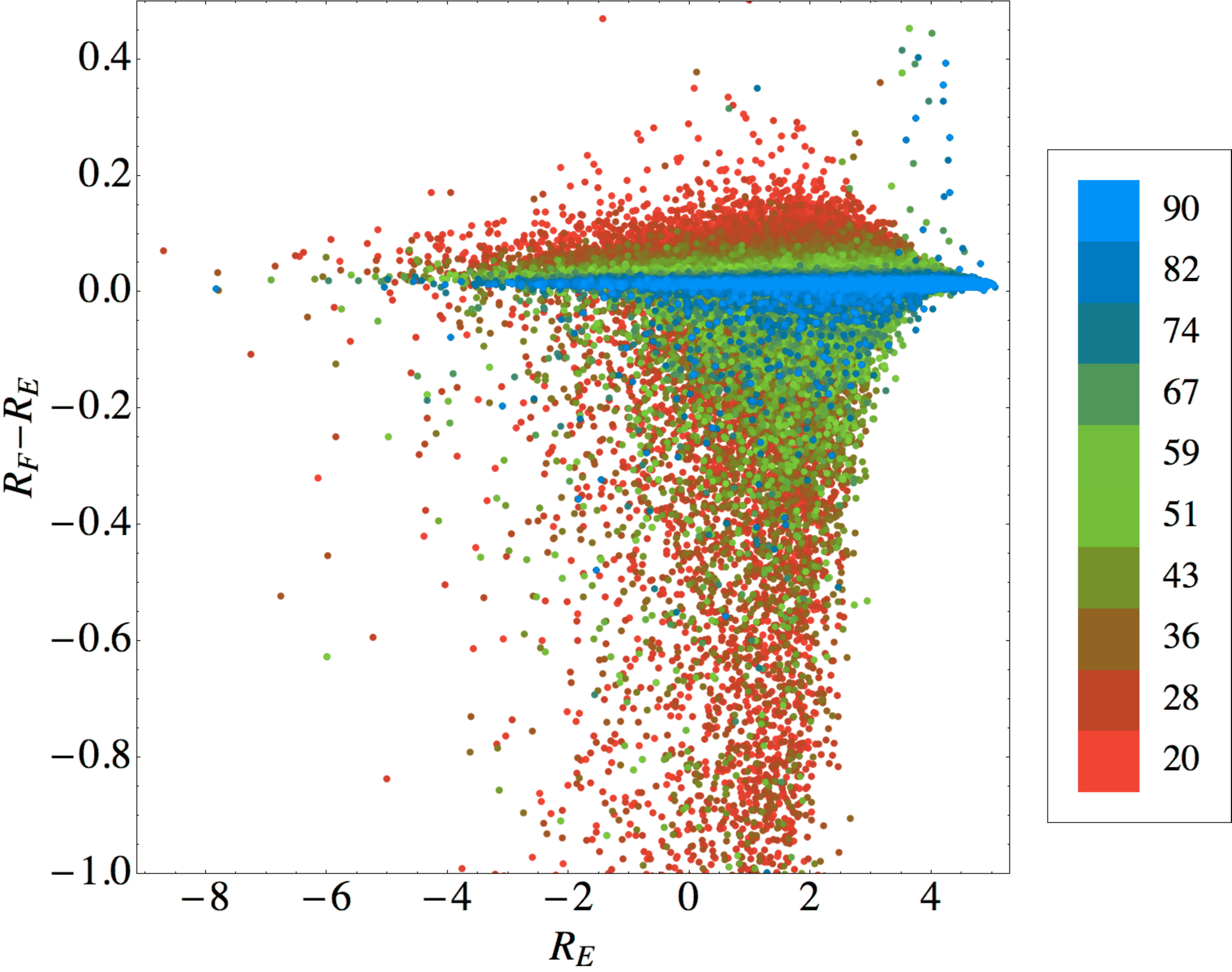}
\caption{Comparison between the exact robustness $R_{E}$ and its Fisher approximation $R_{F}$ for a random set of subsets. Each subset is represented by a point, which is color coded according to the corresponding number $N_{2}$ of SNe in the subset $d_{2}$.
This analysis clearly shows that, as far as the robustness is concerned, the Fisher approximation holds better for larger subset sizes.
See Section \ref{sec:scanna} for more details.}
\label{FIG:CF-CE}
\end{center}
\end{figure}
%%%%%%%%%%%%%%%%%%%%
%%%%%%%%%%%%%%%%%%%%

We will now explain how we actually computed the robustness. For subsets with
$N_{2}$ larger than a certain $N_{2,{\rm med}}$ we have found that the likelihood can be adequately
represented by a Gaussian distribution in the parameter space, so as to
legitimate the Fisher approach. The advantage of using the Fisher matrix is in
the computational speed gain, as only the maximum of the likelihood and few
derivatives have to be found numerically. For smaller subsets $N_{2} \le
N_{2,{\rm med}}$, however, the likelihood deviates from a Gaussian
distribution and we are forced to integrate the likelihood numerically over the
full parameter space. %speed up computation with prior knowledge from fisher
In order to empirically find $N_{2,{\rm med}}$ we have computed both the exact
robustness $R_{E}$ of Eq.~\eqref{hr1} and its Fisher approximation $R_{F}$ of
Eq.~\eqref{eq:robustness} for a random set of subsets. Fig.~\ref{FIG:CF-CE}
shows how the discrepancy decreases as the subset $d_{2}$ becomes larger. As we
will see below in Fig.~\ref{D1-PDF}, the iR-PDF varies on a scale of order unity
in robustness. Therefore, we want to keep the error in the robustness
computation at a level $\lesssim0.1$. We found that the value $N_{2,{\rm
med}}=90$ satisfies this requirement.

%%%%%%%%%%%%%%%%%%%%%%%%%%%%%%%%%%%%
%%%%%%%%%%%%%%%%%%%%%%%%%%%%%%%%%%%%
\section{Results}\label{results}
 %%%%%%%%%%%%%%%%%%%%%%%%%%%%%%%%%%%%
%%%%%%%%%%%%%%%%%%%%%%%%%%%%%%%%%%%%

In analyzing the Union2.1 catalogue of 580 SNe we will restrict to the case of the curved $\Lambda$CDM model, i.e., we allow for spatial curvature but fix the equation of state parameter ($w=-1$). The cosmological parameters are therefore the present-day density parameters $\omegam$ and $\omegal$, with the addition of the nuisance offset $\alpha$. We will consider other parameterizations in forthcoming work.

Our results will be divided into three parts. First in Section \ref{100EdS} we will test our method with a mock dataset whose SNe were drawn from two very different cosmological models. In Section \ref{Union21cuts} we will analyze the Union2.1 dataset~\citep{Suzuki:2011hu} including previously-excluded supernovae (i.e., SNIa that did not pass all the quality selection cuts); this also should be a test for our method. Finally, in Section \ref{Union21} we will present our results regarding the actual Union2.1 dataset.

Before dealing with our results, it is useful to define what we mean by ``mock catalogue''. A mock catalogue $\mock$ for a given dataset $\real$ %with entries $\{ z_{i}, x_{i}, \sigma_{i} \}$
is a synthetic unbiased dataset generated using the best-fit model of $\real$ as the fiducial model. More precisely, we keep fixed the redshifts $z_{i}$ and errors $\sigma_{i}$ of $\real$, and change the distance moduli to $x_{{\rm mock}}=m_{{\rm fiducial}}+x_{{\rm random}}$ where $x_{{\rm random}}$ is drawn from a gaussian distribution of zero mean and standard deviation $\sigma_{i}$.

%The analysis will be similar in the three Sections.

%%%%%%%%%%%%%%%%%%%%%%%%%%%%%%%%%%%%
\subsection{Systematics-driven dataset}
\label{100EdS}
%%%%%%%%%%%%%%%%%%%%%%%%%%%%%%%%%%%%

%%%%%%%%%%%%%%%%%%%%%%%%%%%%%%%%%%%%
\subsubsection{Dataset and iR-PDF computation}
\label{catata}
%%%%%%%%%%%%%%%%%%%%%%%%%%%%%%%%%%%%

\begin{figure}
\begin{center}
\includegraphics[width=\columnwidth]{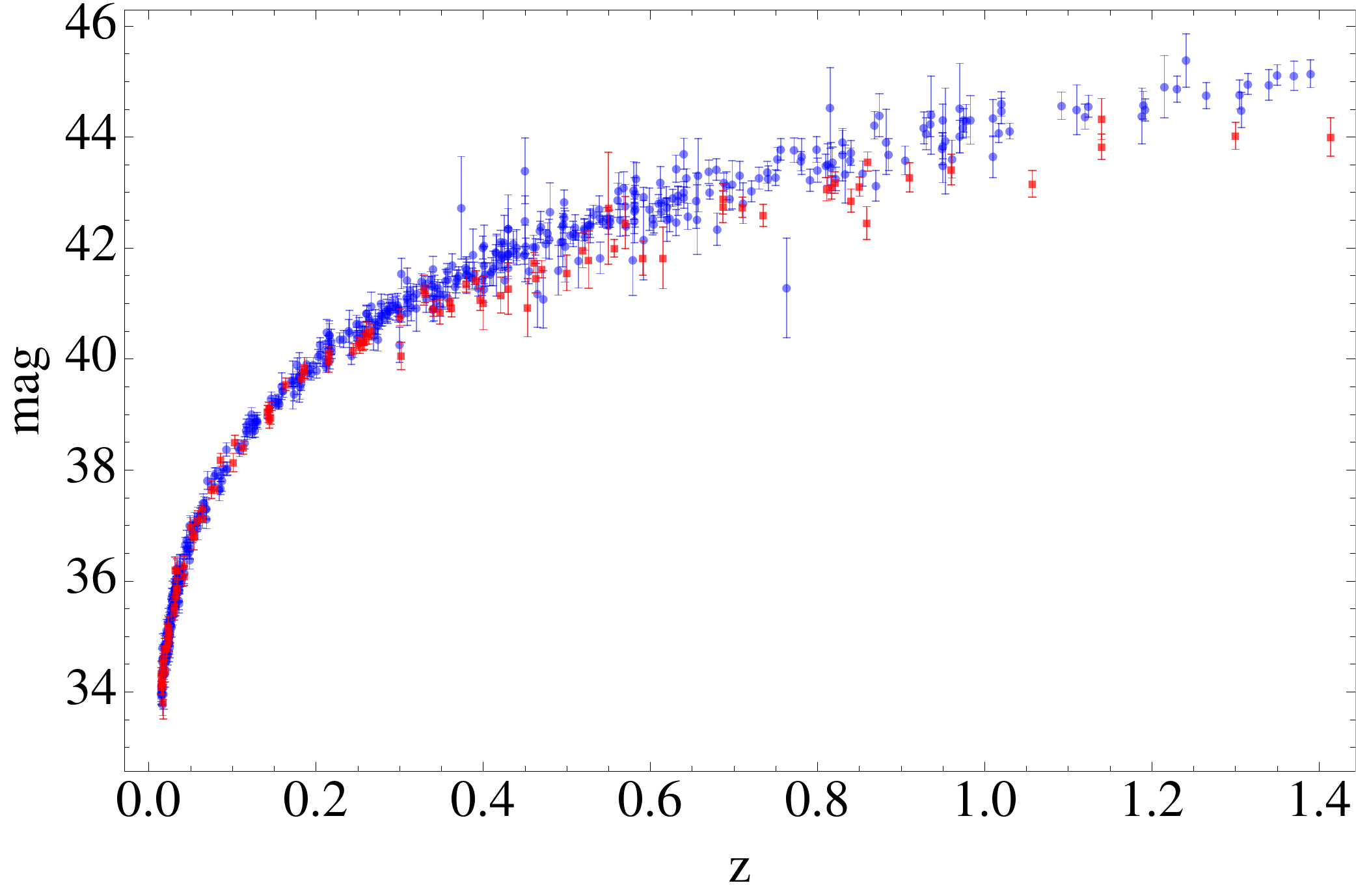}
\caption{Plotted is the distance modulus of the SNe of the systematics-driven dataset $\real_{\rm EdS}$ in which 100 SNe have been drawn from the EdS model as fiducial model (red data) and the remaining from the best-fit model of Union 2.1 (blue data).
Without the different coloring it would be difficult to spot by eye the EdS SNe (the challenged reader could try it on a B\&W printer).
See Section \ref{catata} for more details.}
\label{badset}
\end{center}
\end{figure}

\begin{figure}
\begin{center}
\includegraphics[width=7.5 cm]{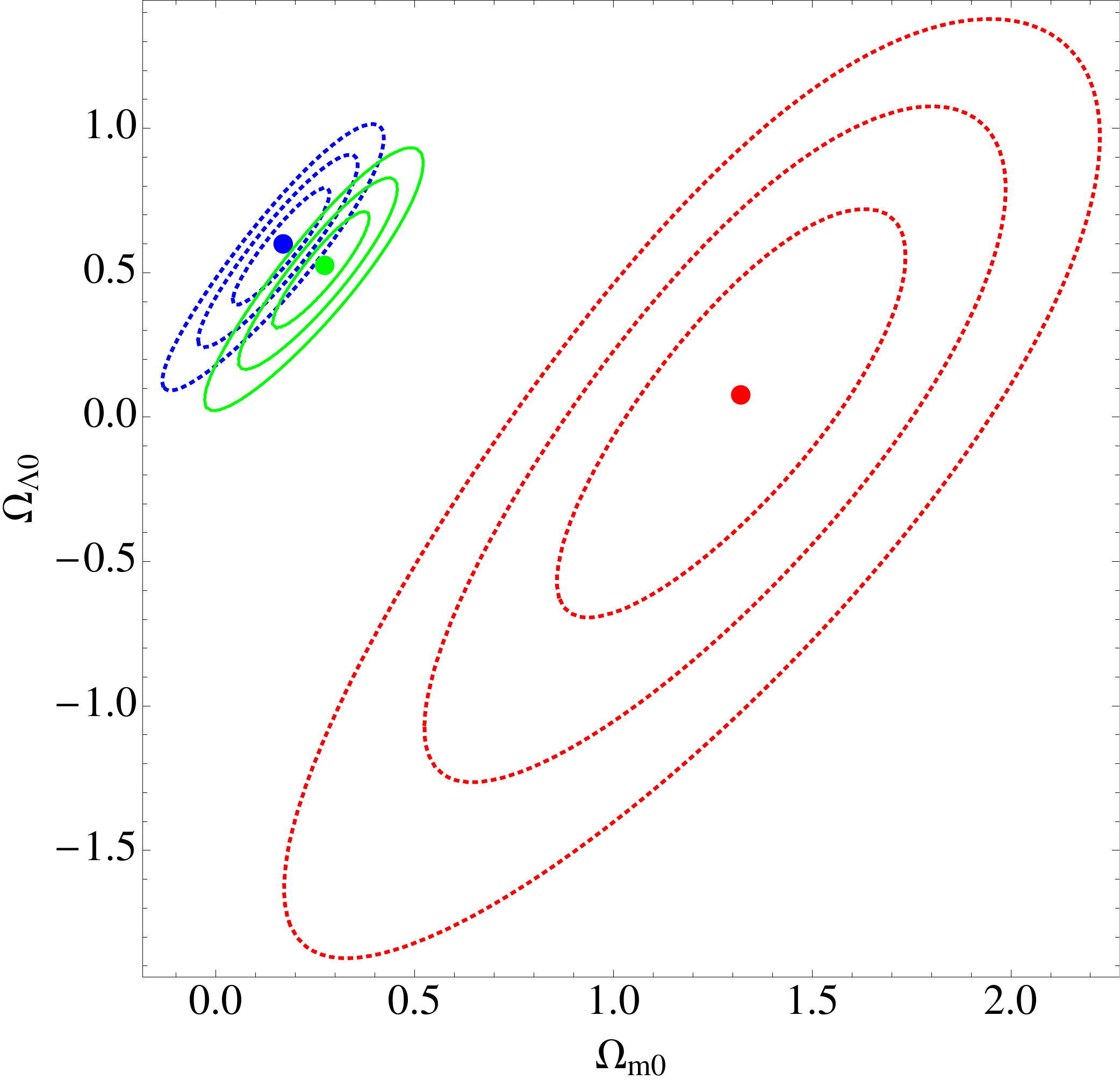}
\caption{1, 2 and 3$\sigma$ confidence-level contours for subsets $d_{1}$ (blue) and $d_{2}$ (red), where $d_{2}$ is made of the 100 systematics-driven EdS SNe shown in red in Fig.~\ref{badset}.
The contours are far apart and the robustness is indeed extremely low, $R \simeq -97.6$.
Also shown in green are the confidence-level contours for the complete dataset $d_{\rm tot}$.
The 100 EdS SNe clearly bias the likelihood surface of the full dataset.
See Section \ref{catata} for more details.}
\label{C-EdS}
\end{center}
\end{figure}

In order to test our method we have generated a systematics-driven dataset $\real_{\rm EdS}$ in the following way. First we have created a mock catalogue of Union2.1. Then we replaced 100 randomly chosen distance-modulus entries with others drawn taking the Einstein-de Sitter (EdS) model (flat, matter-dominated) as the fiducial one (instead of the best-fit model of Union2.1). These 100 SNe are shown in red in Fig.~\ref{badset}. One expects for $\real_{\rm EdS}$ very low robustness values, as for the subset of 100 EdS SNe it is clearly favored the possibility that $d_{1}$ and $d_{2}$ have independent cosmological parameters (which, we remind, we use to parametrize also the systematics). In the case of $d_{2}$ being exactly the subset of 100 EdS SNe, one obtains the plot of Fig.~\ref{C-EdS} which shows the 1, 2 and 3$\sigma$ confidence-level contours for $d_{\rm tot}$ and $d_{1},d_{2}$ independently. The contours are indeed far apart and the robustness is extremely low, $R_{\rm min}\simeq-97.5$ (as previously mentioned, we know \emph{a posteriori} that $R$ is typically a ${\cal O}(1)$ quantity). Fig.~\ref{C-EdS} summarizes nicely the ultimate goal of internal robustness, that is, to go from the original set in green to the ``decontaminated'' set in red. The contours do not look significantly different (i.e.~same precision), but their position has substantially moved by $\sim 3 \sigma$ (i.e.~much better accuracy).

%%%%%%%%%%%%%%%%%%%%
%%%%%%%%%%%%%%%%%%%%
%%%%%%%%%%%%%%%%%%%%
%%%%%%%%%%%%%%%%%%%%

We will now pretend that $\real_{\rm EdS}$ is made of real data and test the sensitivity of our method with it. Before proceeding, however, one can see that the dataset is sick by means of the standard goodness-of-fit test given by $\bar \chi^{2}= \hat \chi^{2}/(N_{\rm tot}-3)$ (the cosmological model has three parameters). For $\real_{\rm EdS}$ we find indeed $\bar \chi^{2}= 1.39$, that is, the catalogue is incompatible with the theoretical model at 6-$\sigma$ level. Nevertheless we advocate that internal robustness is a better test than the standard goodness-of-fit; in other words we would like to show that we can give a stronger exclusion. In order to do so, we will ``normalize'' $\real_{\rm EdS}$ by adding a constant $\sigmaextra$ to the errors $\sigma_{i}$ such that $\bar \chi^{2}= 1$.
This procedure should favor the single cosmological model $M_{C}$ as the normalized catalogue passes the goodness-of-fit test; the idea is to see if the internal-robustness test can still show that the dataset is systematics driven.

%%%%%%%%%%%%%%%%%%%%
%%%%%%%%%%%%%%%%%%%%
%%%%%%%%%%%%%%%%%%%%
%%%%%%%%%%%%%%%%%%%%

The technique of normalizing a catalogue to $\bar \chi^{2}= 1$ is actually often used in supernova cosmology~\citep{Astier:2005qq}. SNe are indeed imperfect standard candles with a residual scatter, called the intrinsic dispersion, of roughly $\sigma_{\rm int}\sim 0.1$ magnitudes. As SN physics is not yet thoroughly understood, $\sigma_{\rm int}$ is not tightly constrained, and it is often determined by demanding that $\bar \chi^{2}= 1$.\footnote{Note that this procedure may hide problems with the theoretical model being used, as it is shown by the very example of~$\real_{\rm EdS}$. For further discussion and alternatives see~\citealt{Kim:2011hg,March:2011xa,Lago:2011pk}.}
SNe catalogues are therefore perfect candidates for testing the internal robustness method. Since the Union2.1 data already include implicitly a $\sigma_{\rm int}$, what we call $\sigmaextra$ is the amount to be added to $\sigma_{\rm int}$ in order to have $\bar \chi^{2}= 1$. For $\real_{\rm EdS}$ one needs $\sigmaextra =0.0356$ magnitudes; that is, to increase the errors. As a consequence the contours of Fig.~\ref{C-EdS} become slightly broader and the minimum robustness is larger, even though still extremely low: $R_{\rm min}\simeq-70.9$.

Finally, we computed the internal robustness for the (normalized) $\real_{\rm EdS}$ dataset using the set of partitions $\Xi(T)$ with a statistics of $T=5\cdot10^{5}$ subsets (see Section~\ref{sec:scanna}). As explained in Section~\ref{spir}, in order to determine if the iR-PDF of $\real_{\rm EdS}$ passes or not the robustness test, we have to generate a distribution of iR-PDFs, which we obtain by computing the internal robustness for 100 mocks $\mock_{j}$. Each mock iR-PDF is generated using the same set prescription $\Xi(T)$ but with a lower statistics of $T=5\cdot10^{4}$, as the fluctuations among the mocks are more important than the ``sampling'' fluctuations due to poissonian errors. The mock catalogues $\mock_{j}$ have been also normalized to $\bar \chi^{2}= 1$. It is worth saying at this point that the robustness test is quite expensive from a computational point of view. The numerical results of this paper have been obtained with Wolfram Mathematica 8 and the average CPU time to calculate the robustness value of a given partition was $\sim 2-3$ seconds (luckily the computation is easily parallelized). The size of $T$ and the number of mocks used are therefore the main constraints in the final results: the higher the statistics, the clearer the signal one may get.

%%%%%%%%%%%%%%%%%%%%
%%%%%%%%%%%%%%%%%%%%
\begin{figure*}
\begin{center}
    \includegraphics[height= 6.05 cm]{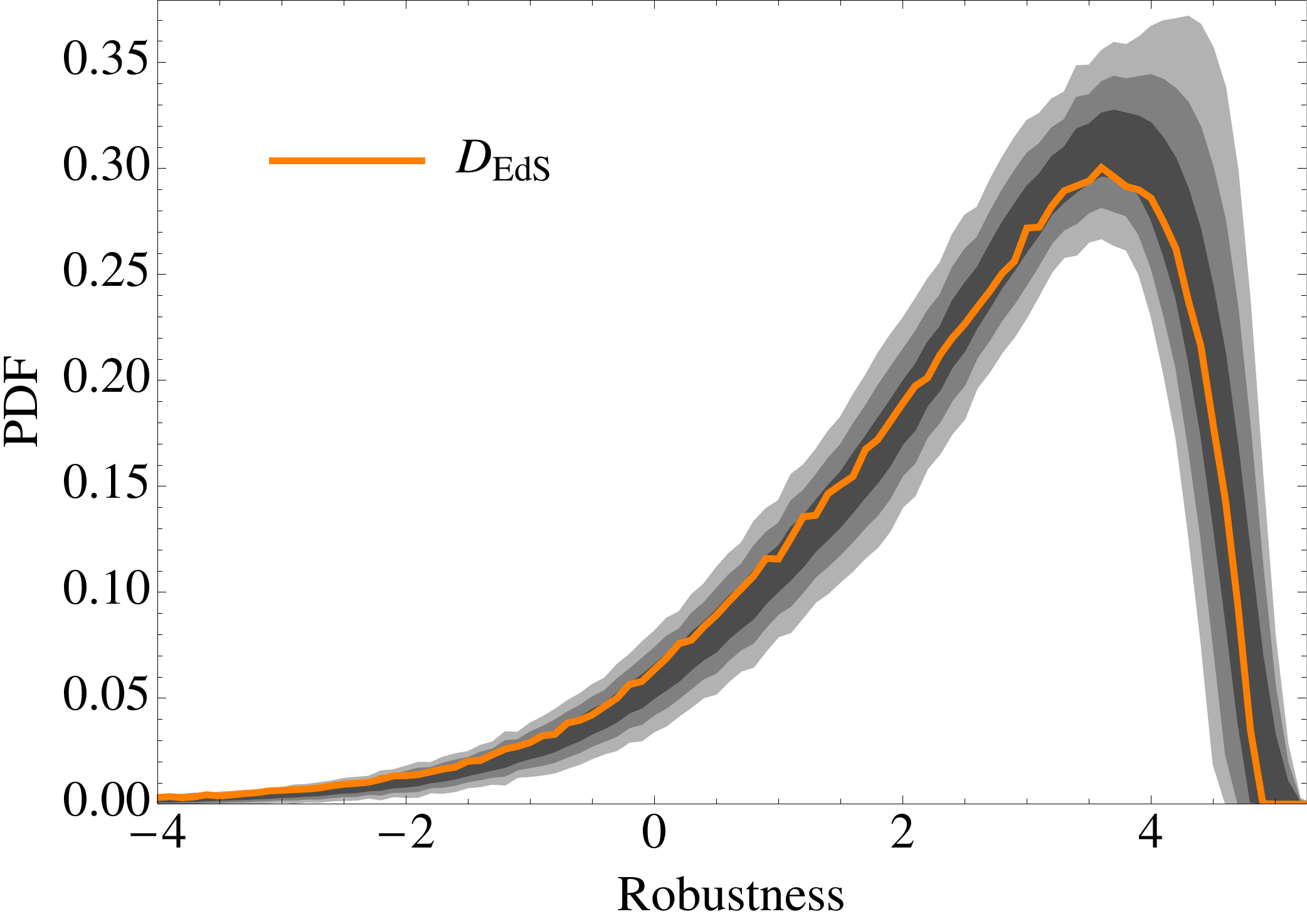}
    \qquad
    \includegraphics[height= 6.05 cm]{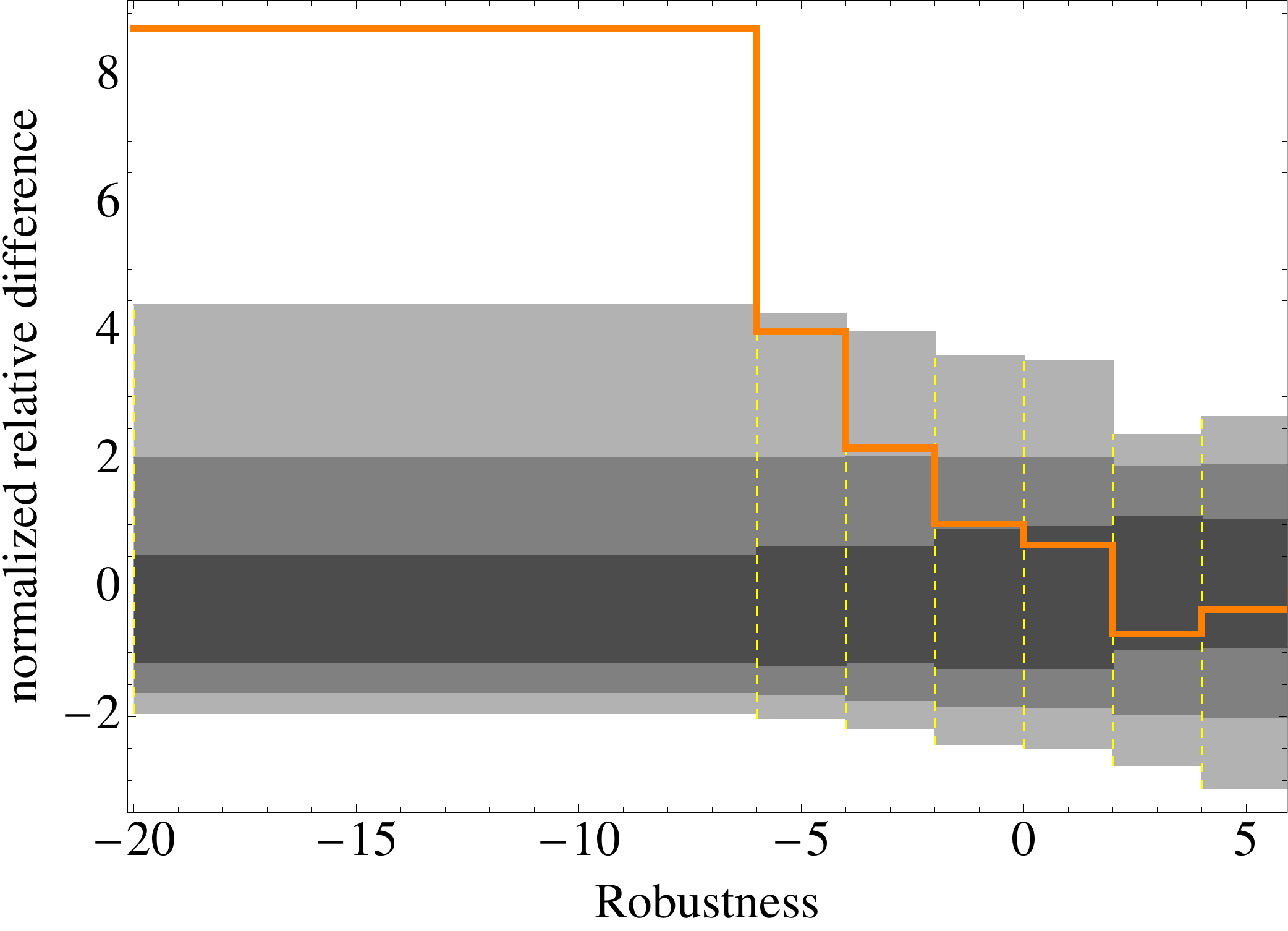}
    \caption{[Left]: binned internal-robustness PDF (orange curve) for the dataset $\real_{\rm EdS}$ of Fig.~\ref{badset} against $\sigma$-bands (gray areas) from (unbiased) mock catalogues.
    [Right]: same as the left panel for larger bins (dashed lines) and lower robustness values. Moreover, the bin height values $h_{k}$ have been translated and scaled according to $h_{k} \rightarrow (h_{k} - \bar h_{k})/  \sigma_{h_{k}}$ in order to uniformly show the signal across the various bins.  One can clearly see that while the body of the iR-PDF is compatible with the mocks, the low-robustness tail is detected as being driven by systematics. The $\real_{\rm EdS}$ datum relative to the bin $(-20,-6]$ lies at 4.2-$\sigma$ confidence levels. See Section~\ref{anapdf} for more details.}
    \label{D1-PDF}
\end{center}
\end{figure*}
%%%%%%%%%%%%%%%%%%%%%
%%%%%%%%%%%%%%%%%%%%%

%%%%%%%%%%%%%%%%%%%%%%%%%%%%%%%%%%%%
\subsubsection{Analysis of the iR-PDF}
\label{anapdf}
%%%%%%%%%%%%%%%%%%%%%%%%%%%%%%%%%%%%

The left panel of Fig.~\ref{D1-PDF} shows the iR-PDF of $\real_{\rm EdS}$ (solid orange line), which has been obtained by binning the robustness values within $N_{b}$ bins of widths $\Delta R_{k}$.
The same binning prescription has been used to calculate also the bin heights $h_{kj}$ of the unbiased mocks $\mock_{j}$, which have been used to compute mean $\bar h_{k}$ and standard deviation $\sigma_{h_{k}}$ for the $N_{b}$ bins.
By assuming that the bin heights within a given bin are distributed according to a gaussian PDF with mean $\bar h_{k}$ and the standard deviation $\sigma_{h_{k}}$, we have then drawn the $\sigma$-bands bounding a systematics-free iR-PDF (gray areas in the left panel of Fig.~\ref{D1-PDF}).
From this plot it seems as if the iR-PDF of $\real_{\rm EdS}$ passes the internal robustness test.

One expects, however, the signal to be concentrated in the low-robustness
tail of the PDF, in which systematics-driven SNe should dominate one of the two partitions $d_{1,2}$.
The spurious iR-PDF can indeed be loosely thought as the sum of a systematics-free PDF and a systematics-driven
perturbation.
The systematics-free PDF will behave according to the $\sigma$-bands, which do not change sizably if the biased subset is sufficiently small.
The perturbation will then be given by the 100 EdS SNe which, when $d_{2}$ coincides with them, will add a {\em
single} point to the histogram at $R_{\rm min}\simeq-70.9$. This point is
clearly not observable in practice as one cannot scan all the $10^{174}$
possible subsets. At the expense of a higher robustness, however,
there will be many subsets $d_{2}$ with a fraction of the 100 EdS SNe, thus generating a stronger low-robustness
tail in a biased dataset.

\begin{figure*}
\begin{center}
    \includegraphics[height= 6.3 cm]{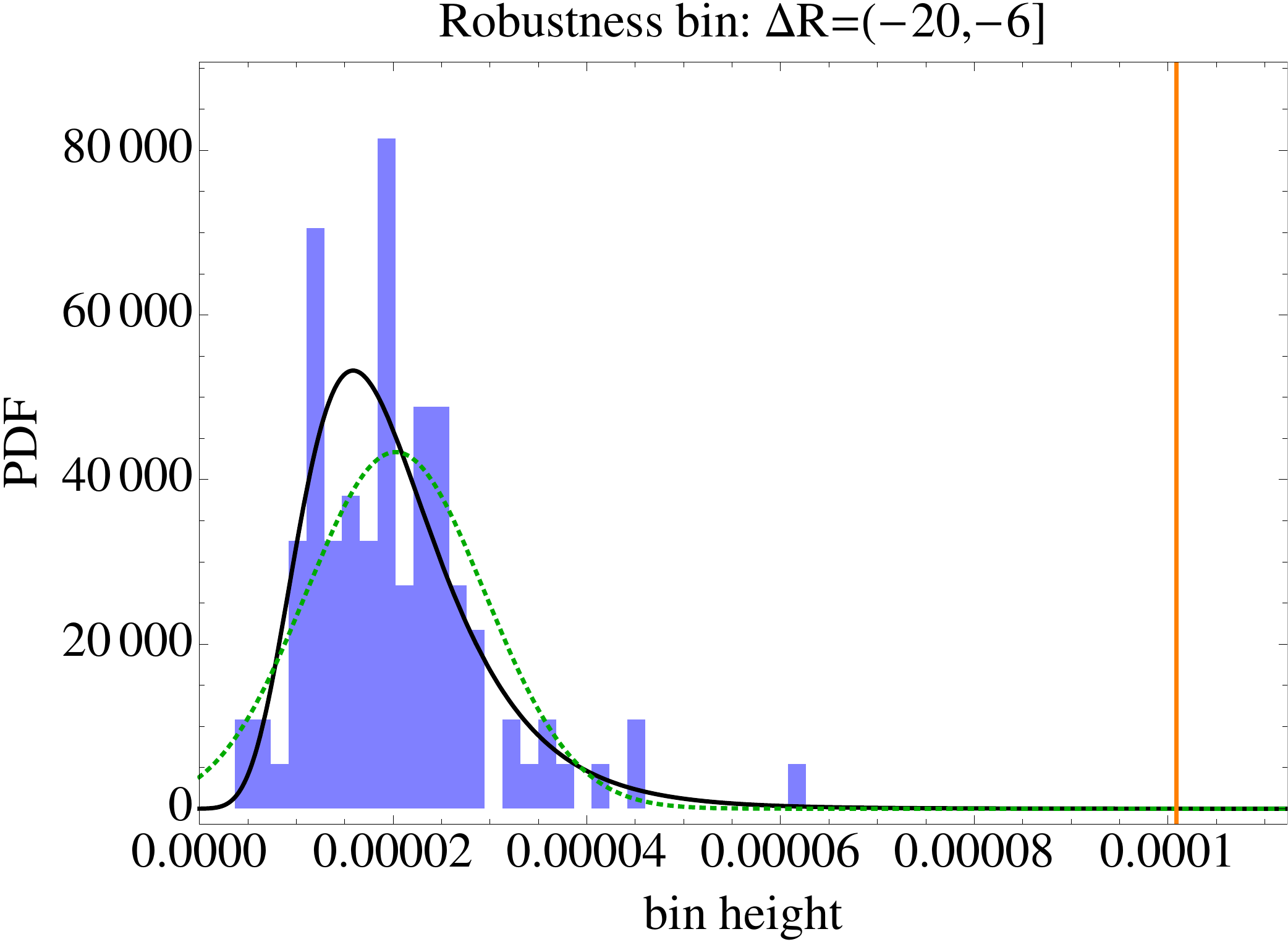}
    \qquad
    \includegraphics[height= 6.3 cm]{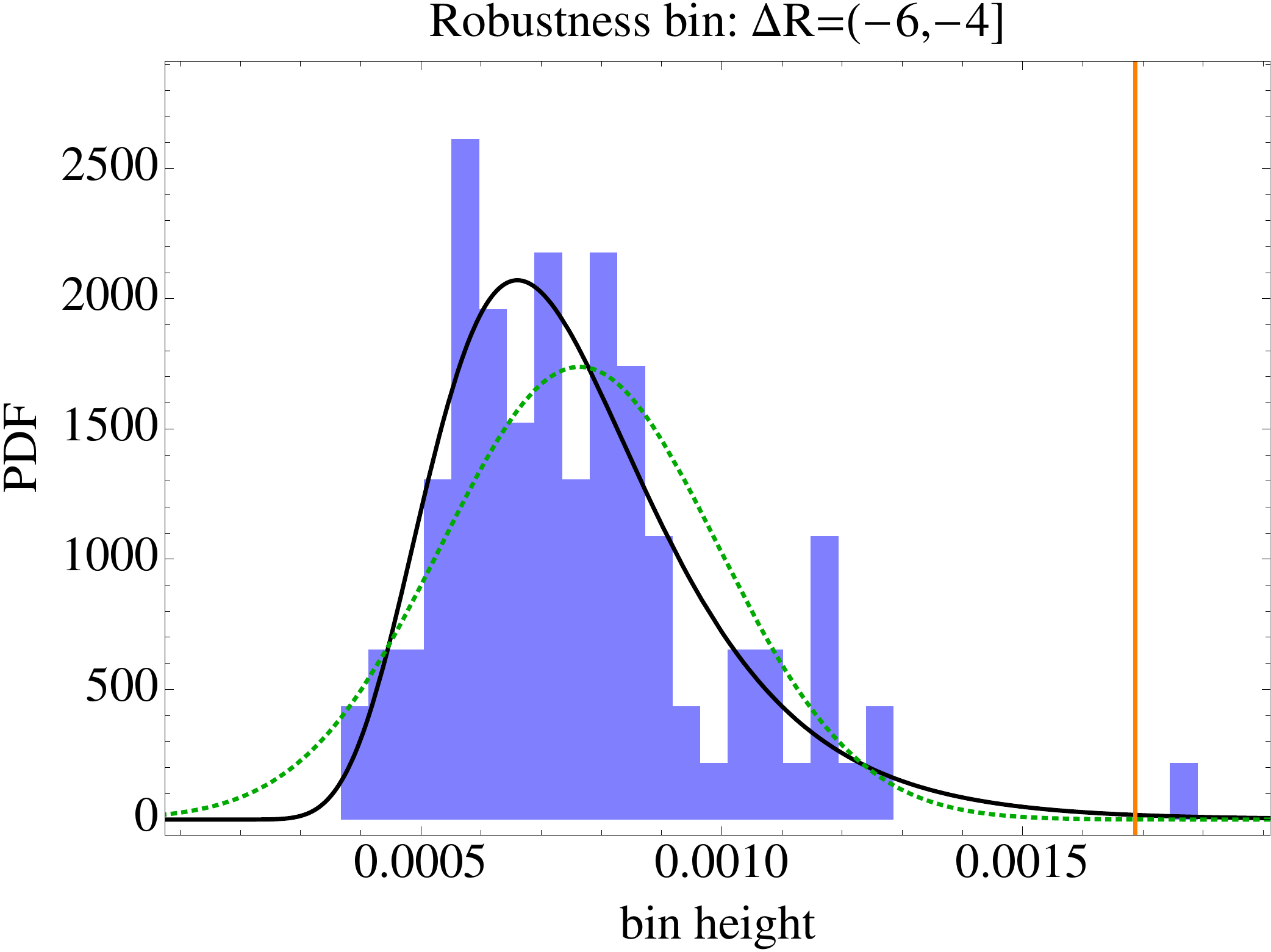}
    \caption{Binned distributions of bin heights for the lowest (left panel) and next-to-lowest (right panel) robustness bins of the mock robustness distributions used to make the $\sigma$-bands shown on the right panel of Fig.~\ref{D1-PDF}.  The bin height corresponding to $\real_{\rm EdS}$ is shown as an orange vertical line, and corresponds to the orange curve on the right panel of Fig.~\ref{D1-PDF}. The dotted line is a fitted gaussian distribution. The solid line is the fitted Pearson distribution which has been used to draw the $\sigma$-bands shown on the right panel of Fig.~\ref{D1-PDF}. See Section \ref{anapdf} for more details.}
    \label{D1-bins}
\end{center}
\end{figure*}

In order to analyze the low-robustness tail of the PDF, we have to drop the gaussian assumption.
The latter is indeed a good approximation only for the bins in the body for which $\bar h_{k}/\sigma_{h_{k}}\gg1$, but not in the tail as illustrated in Fig.~\ref{D1-bins} where the distributions for two low-robustness bins are shown.
The reason is that the iR-PDFs of the mocks have a compact support which have a variable $R_{\rm min}$. At a given low value $R$ in the tail, therefore, some of the robustness distributions of the mocks will be very close (if not identical) to zero, with the consequence that the distribution of the bin heights will be skewed, thus deviating from gaussianity.
Note that this cannot be cured by using a higher statistics $T$ as this feature is related to the existence of an $R_{\rm min}$ for the iR-PDF.

In order to analyze the tail, the standard way to proceed would be to generate many mock iR-PDFs so as to compute the non-gaussian distribution in the bin heights numerically.
As remarked earlier, however, it is numerically expensive to obtain an iR-PDF and we limit our sample of mock catalogues to 100.
In order to properly quantify the signal we have then proceeded in two ways.
The first and most obvious is to check if any mock has a bin height larger than $\real_{\rm EdS}$. As shown by Fig.~\ref{D1-bins} this is not the case for the lowest bin (where we expect the signal to be strongest), and we can so conclude that $\real_{\rm EdS}$ is systematics-driven at a confidence level better than 99\%, or 2.6$\sigma$.

%%%%%%%%%%%%%%%%%%%%
%%%%%%%%%%%%%%%%%%%%
%%%%%%%%%%%%%%%%%%%%%
%%%%%%%%%%%%%%%%%%%%%

The second way is to use a template to be fitted to the data, and then use the fitted template to assess the significance of the datum relative to $\real_{\rm EdS}$. We used as template the Pearson distribution~\citep{1895RSPTA.186..343P,1916RSPTA.216..429P}, which is found by demanding that its first four moments coincide with the moments from the bin heights. The result is depicted by a solid black in Fig.~\ref{D1-bins}, where also the gaussian distribution is plotted for comparison. As one can see the fitted Pearson distribution correctly reproduces the skewness of the data, and in particular does not go to negative bin height, as instead does the gaussian distribution for the lowest bin.

Having calculated the $\sigma$-levels with the Pearson template, we can now show our final results in the right panel of Fig.~\ref{D1-PDF}.
We have used larger bins as compared to the plot in the left panel because the bins extend to lower robustness values, which have lower statistics.
Also, the bin height values $h_{k}$ have been translated and scaled according to:
\begin{equation}
    h_{k} \longrightarrow {h_{k} - \bar h_{k} \over \sigma_{h_{k}}} \,,
\end{equation}
so as to uniformly show the signal across the various bins.
Fig.~\ref{D1-PDF} clearly shows how the body of the iR-PDF of $\real_{\rm EdS}$ passes the robustness test, as opposed to the low-robustness tail which is detected as being systematics-driven.
More precisely, the $\real_{\rm EdS}$ datum relative to the bin $(-20,-6]$ lies at 4.2-$\sigma$ confidence levels, while the datum relative to the robustness bin $(-6,-4]$ lies at 2.8-$\sigma$ confidence levels.
Note that in the previous results we did not include the error in the datum relative to $\real_{\rm EdS}$. The latter is indeed negligible as the iR-PDF of $\real_{\rm EdS}$ has been found with a statistics much higher than the one relative to the iR-PDF of the mocks.

Finally, one would expect non-negligible fluctuations in the third (skewness) and forth (kurtosis) moments coming from a sample of only 100 data, and in order to assess the uncertainty on the exclusion signal, one may proceed as follows.
Repeat enough times: a) generate 100 random values from the fitted Pearson PDF; b) fit again the Pearson PDF to this new data and calculate the $\sigma$ confidence level.
If we now apply this routine to the two lowest bins of Fig.~\ref{D1-bins} for which the signal is 4.2$\sigma$ and $2.8\sigma$, respectively, we find that the signal is $>3.3\sigma$ and $>2.3\sigma$ at 95\% confidence level, respectively.

%%%%%%%%%%%%%%%%%%%%%%%%%%%%%%%%%%%%
\subsubsection{Taking correlations into account}
\label{correse}
%%%%%%%%%%%%%%%%%%%%%%%%%%%%%%%%%%%%

%%%%%%%%%%%%%%%%%%%%
%%%%%%%%%%%%%%%%%%%%
\begin{figure}
\begin{center}
    \includegraphics[width=\columnwidth]{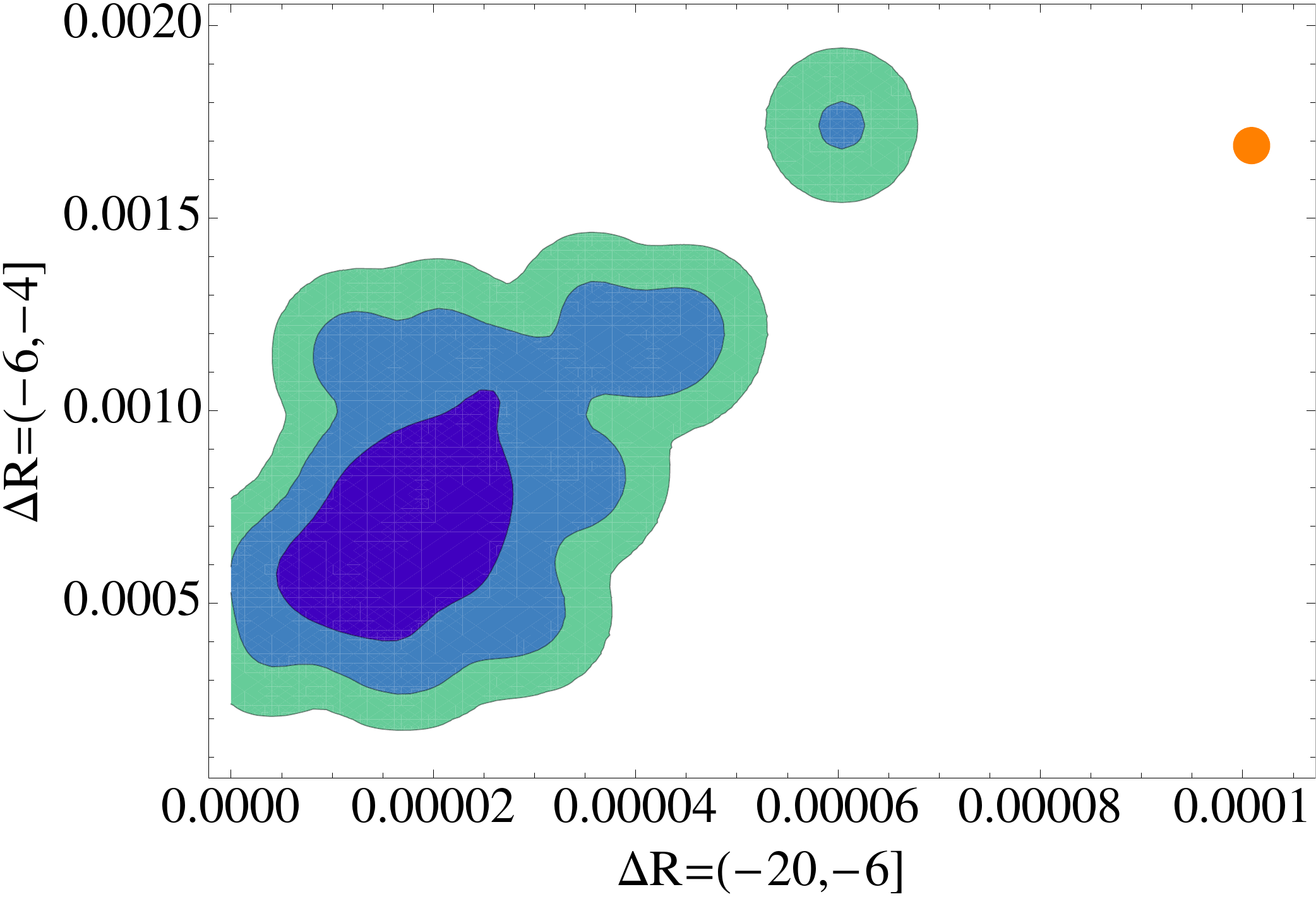}
    \caption{Correlated distribution of bin heights for the two lowest-robustness bins of Fig.~\ref{D1-bins}.
The 1, 2 and 3$\sigma$ confidence-level contours have been calculated using the gaussian values $\Delta \chi^{2}=2.30,6.17, 11.8$.
The bin height corresponding to $\real_{\rm EdS}$ is shown as an orange disk.
See Section \ref{correse} for details.}
    \label{D1-corre}
\end{center}
\end{figure}
%%%%%%%%%%%%%%%%%%%%
%%%%%%%%%%%%%%%%%%%%

When interpreting the results in Fig.~\ref{D1-PDF} one cannot neglect the correlations among the bins $\Delta R_{k}$. In other words, to get a reliable estimate by eye one is supposed to consider only one bin at a time (namely the bin with the strongest signal) and is not allowed to combine the signals from various bins as (all) the bins are most likely correlated. The same procedure can also be followed numerically to get the simplest conservative estimate (see previous Section).

As the signal is concentrated in the low-robustness tail, where the gaussian assumption is not a good approximation, we compute the correlation by building a histogram.
In Fig.~\ref{D1-corre} we show the corresponding contour plot for the two lowest-robustness bins together with the corresponding datum of $\real_{\rm EdS}$ (orange disk). As with only 100 mocks it is not possible to obtain a good enough PDF which can then be integrated to calculate the contours, in Fig.~\ref{D1-corre} the contours are drawn using gaussian values, to wit: $\Delta \chi^{2}=2.30, \,6.17, \,11.8$. From this analysis it looks as if taking correlations into consideration would increase the signal.

%%%%%%%%%%%%%%%%%%%%%%%%%%%%%%%%%%%%
\subsubsection{Frequentist limit}
\label{limichi}
%%%%%%%%%%%%%%%%%%%%%%%%%%%%%%%%%%%%

%%%%%%%%%%%%%%%%%%%%
%%%%%%%%%%%%%%%%%%%%
\begin{figure}
\begin{center}
    \includegraphics[width=\columnwidth]{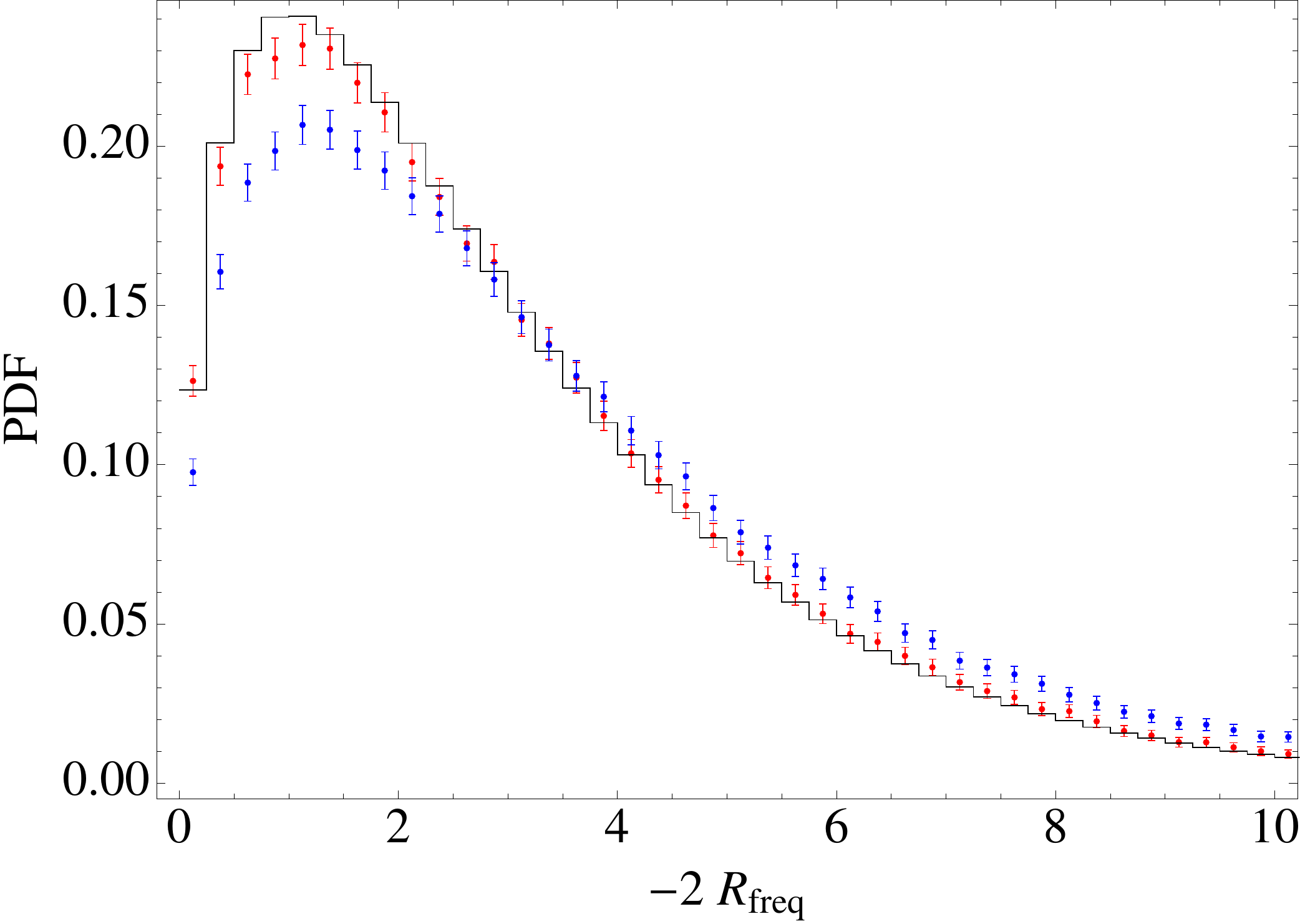}
    \caption{Binned iR-PDF using for the robustness its frequentist limit $-2 R_{\rm freq} \equiv \hat \chi_{{\rm tot}}^{2}-\hat \chi_{1}^{2}-\hat \chi_{2}^{2}$.
    Distributions relative to two unbiased mock catalogues are shown in red and blue, while the binned $\chi^{2}$ distribution with 3 d.o.f. is shown with a black curve. The error bars stand for $3\sigma$. It is clear that the $\chi^{2}$ distribution with 3 d.o.f. is not the correct PDF even though it captures the overall shape.
    See Section \ref{limichi} for more details.}
    \label{D1-freq}
\end{center}
\end{figure}
%%%%%%%%%%%%%%%%%%%%
%%%%%%%%%%%%%%%%%%%%

If one neglects the logarithmic part in Eq.~(\ref{eq:robustnessMa}), then the robustness for the unbiased mocks becomes the parameter goodness-of-fit test \citep{Maltoni:2002xd,Maltoni:2003cu} which is distributed as a $\chi^{2}$ distribution with 3 degrees of freedom.
As explained in Section \ref{spir}, however, this is exactly true only for the external robustness, but not for the internal robustness.
Nevertheless, the $\chi^{2}$ distribution with 3 d.o.f. captures the overall behavior of a fiducial iR-PDF, as can be seen in Fig.~\ref{D1-freq} where is plotted the quantity
\begin{equation}
    R_{\rm freq} \,\equiv\, - {1\over 2} \left( \hat \chi_{{\rm tot}}^{2}-\hat \chi_{1}^{2}-\hat \chi_{2}^{2} \right)
\end{equation}
for two mock catalogues (red and blue) together with the binned $\chi^{2}$ distribution with 3 d.o.f. (black). Note that when the best fits of $d_{1}$ and $d_{2}$ coincide one has $R_{\rm freq}=0$.

%%%%%%%%%%%%%%%%%%%%%%%%%%%%%%%%%%%%
\subsection{Union2.1 dataset with previously-excluded SNe}
\label{Union21cuts}
%%%%%%%%%%%%%%%%%%%%%%%%%%%%%%%%%%%%

%%%%%%%%%%%%%%%%%%%%
%%%%%%%%%%%%%%%%%%%%
\begin{figure}
\begin{center}
    \includegraphics[width=\columnwidth]{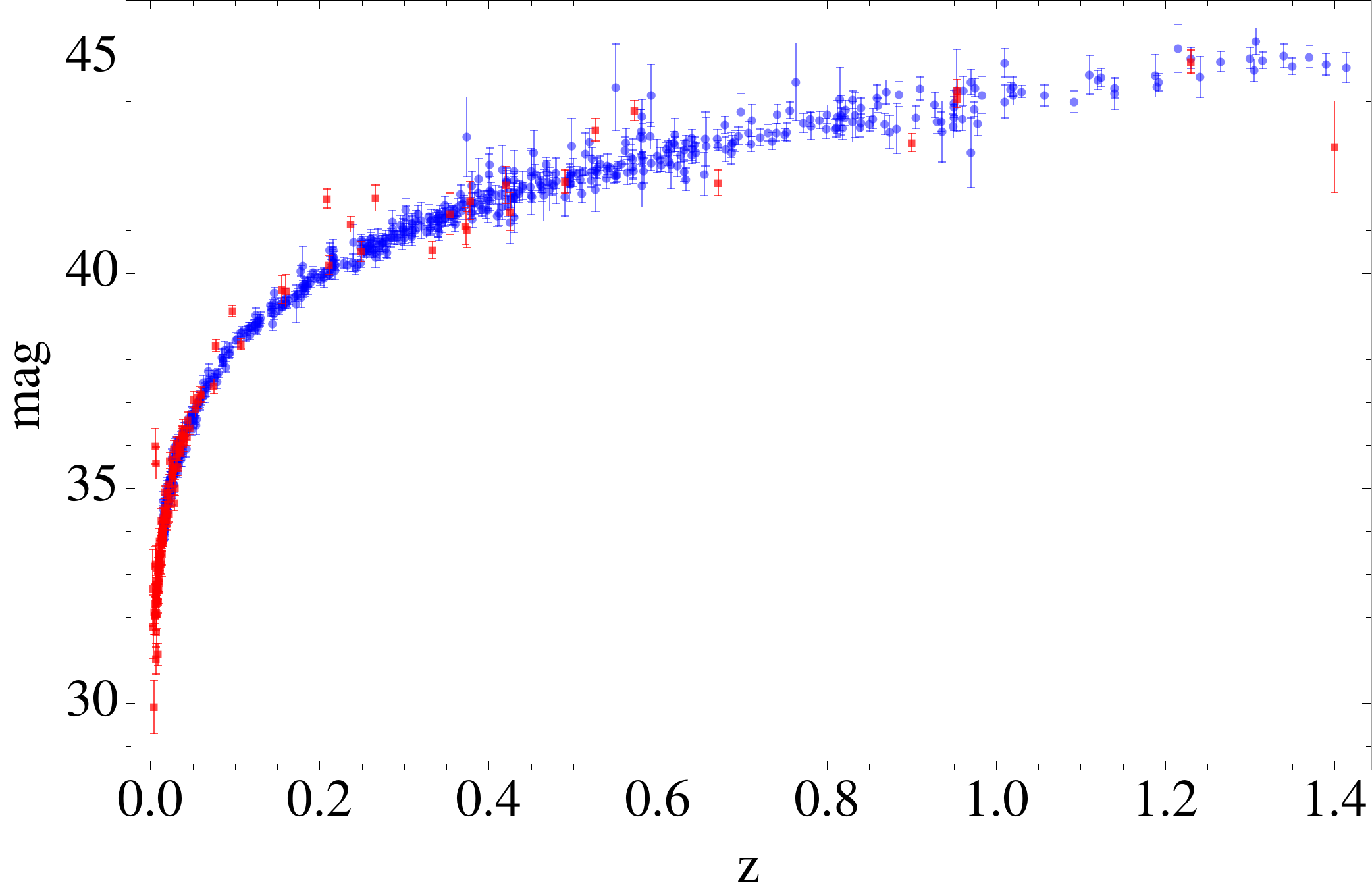}
    \caption{Plotted is the distance modulus of the SNe that passed the quality cuts (blue) and made it into the final Union 2.1 catalogue, and the distance modulus of the SNe that did not (red).
    See Section \ref{Union21cuts} for more details.}
    \label{cuts}
\end{center}
\end{figure}
%%%%%%%%%%%%%%%%%%%%
%%%%%%%%%%%%%%%%%%%%

%%%%%%%%%%%%%%%%%%%%
%%%%%%%%%%%%%%%%%%%%
\begin{figure*}
\begin{center}
    \includegraphics[height= 6.3 cm]{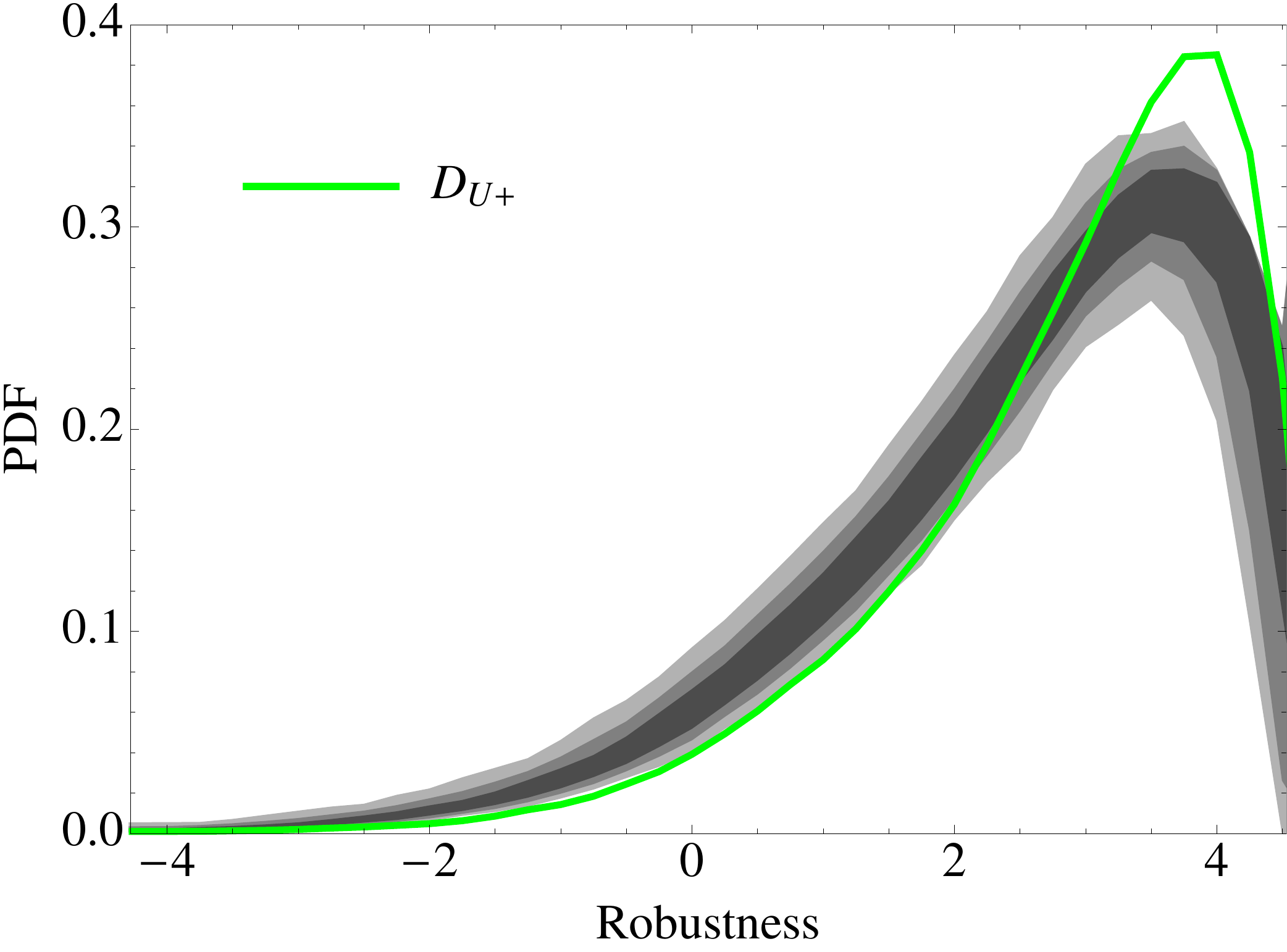}
    \qquad
    \includegraphics[height= 6.3 cm]{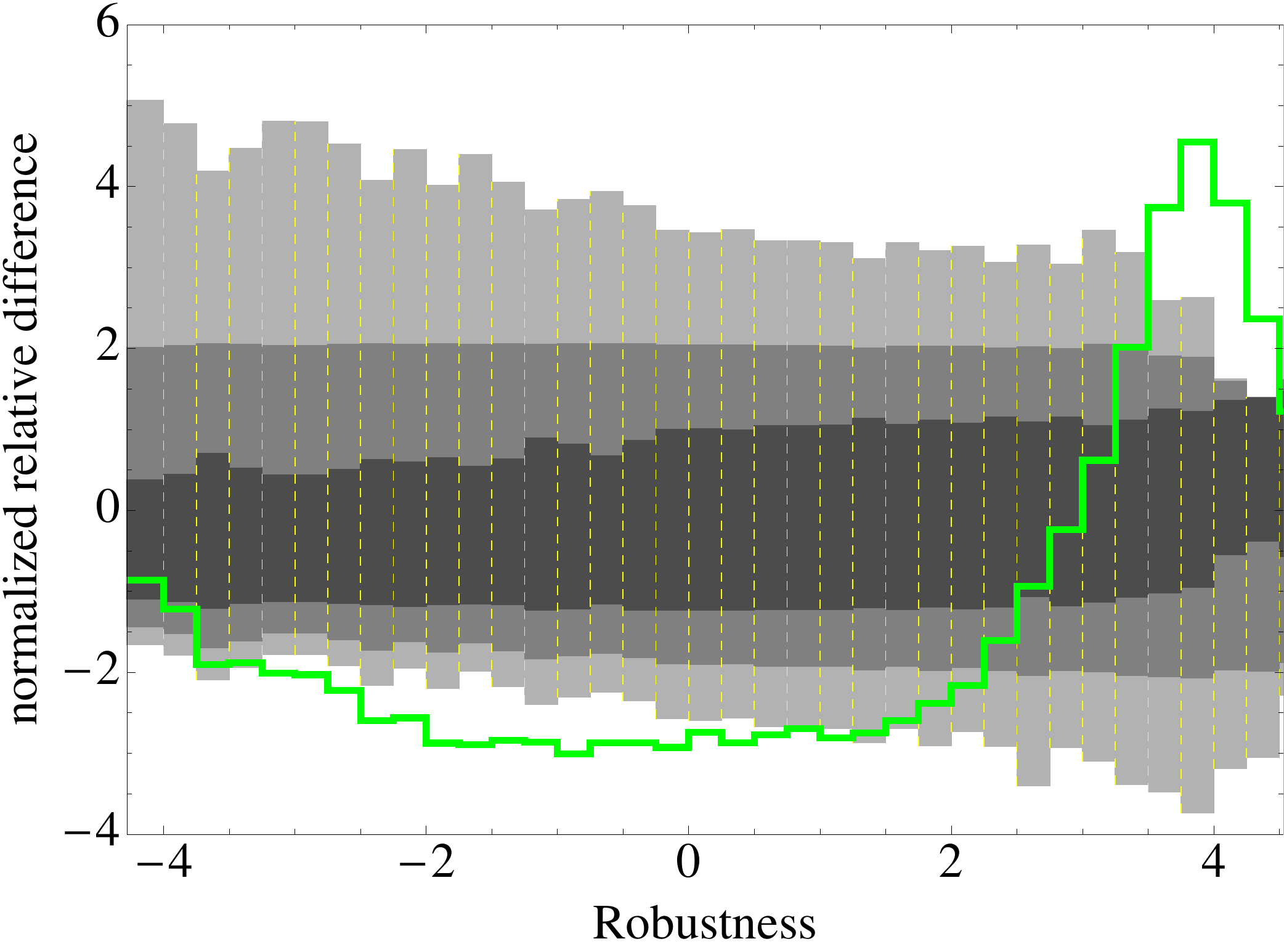}
    \caption{ Same as Fig.~\ref{D1-PDF} for the dataset $\real_{\rm U+}$ of Fig.~\ref{cuts} (green curve). One can clearly see that in this case the (strong) signal of $\real_{\rm U+}$ as being systematics-driven comes from the bins in the body of the PDF.  See Section \ref{Union21cuts} for more details.}
    \label{D2-PDF}
\end{center}
\end{figure*}
%%%%%%%%%%%%%%%%%%%%%
%%%%%%%%%%%%%%%%%%%%%

%%%%%%%%%%%%%%%%%%%%
%%%%%%%%%%%%%%%%%%%%
\begin{figure*}
\begin{center}
    \includegraphics[height= 6.5 cm]{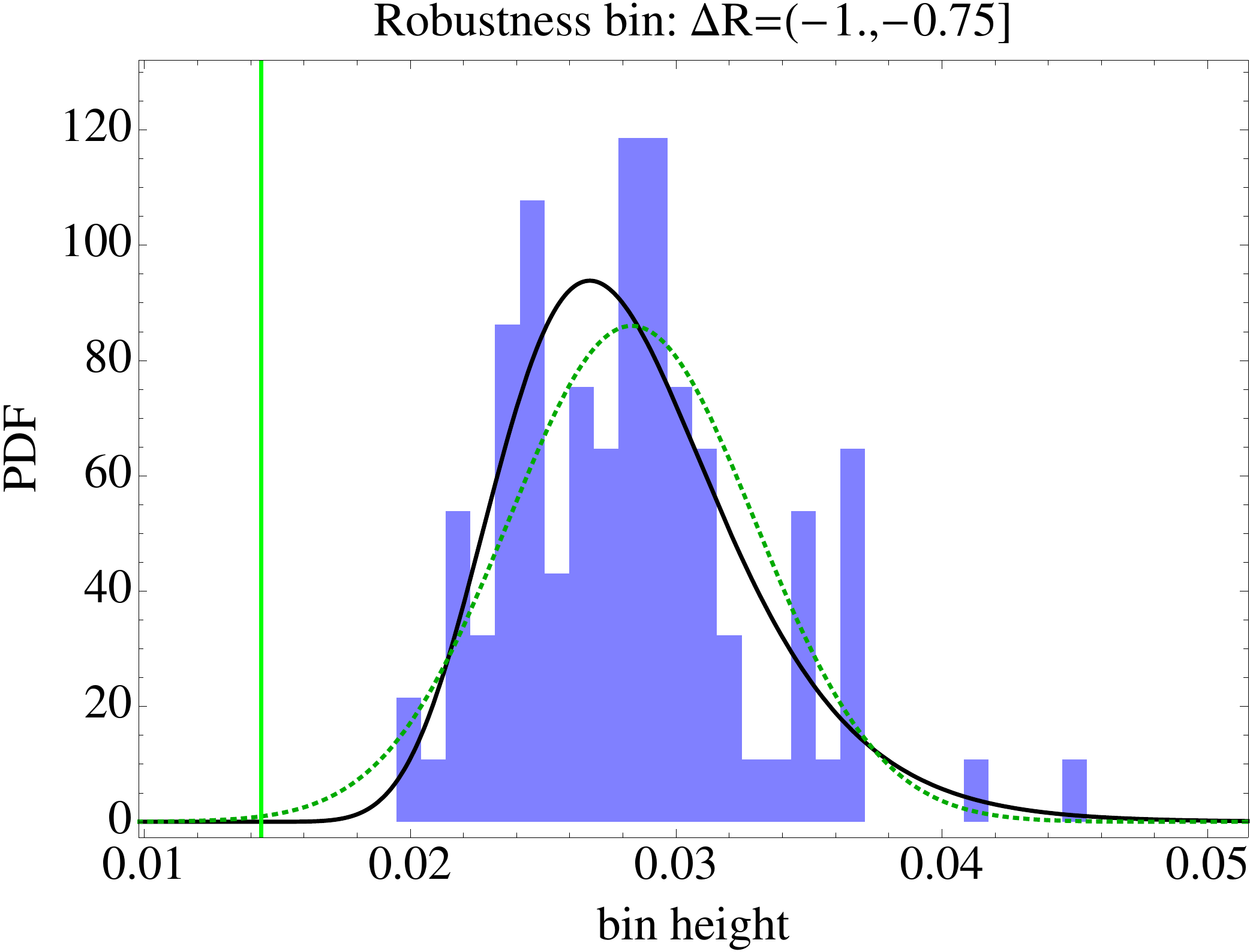}
    \qquad
    \includegraphics[height= 6.5 cm]{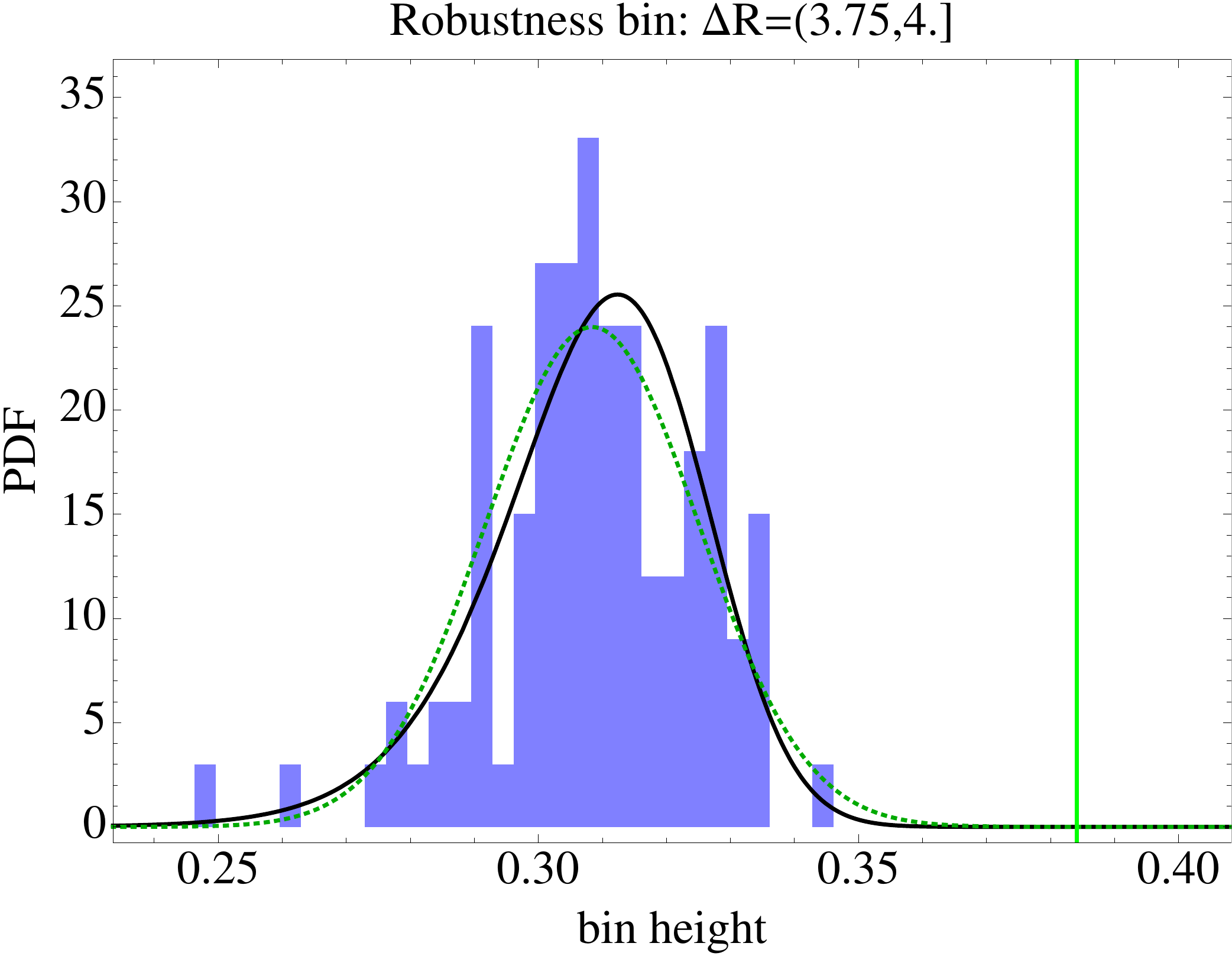}
    \caption{Same as Fig.~\ref{D1-bins} for the dataset $\real_{\rm U+}$. The bin height corresponding to $\real_{\rm U+}$ is shown as a green vertical line, and corresponds to the green curve in Fig.~\ref{D2-PDF}. These two robustness bins were chosen for this plot as they quite strongly show that $\real_{\rm U+}$ is systematics driven.
More precisely, the $\real_{\rm U+}$ datum relative to the bin $(3.75,4]$ lies at 5.4-$\sigma$ confidence levels.
See Section \ref{Union21cuts} for more details.}
    \label{D2-bins}
\end{center}
\end{figure*}
%%%%%%%%%%%%%%%%%%%%%
%%%%%%%%%%%%%%%%%%%%%

%%%%%%%%%%%%%%%%%%%%
%%%%%%%%%%%%%%%%%%%%
\begin{figure}[h!]
\begin{center}
    \includegraphics[width=\columnwidth]{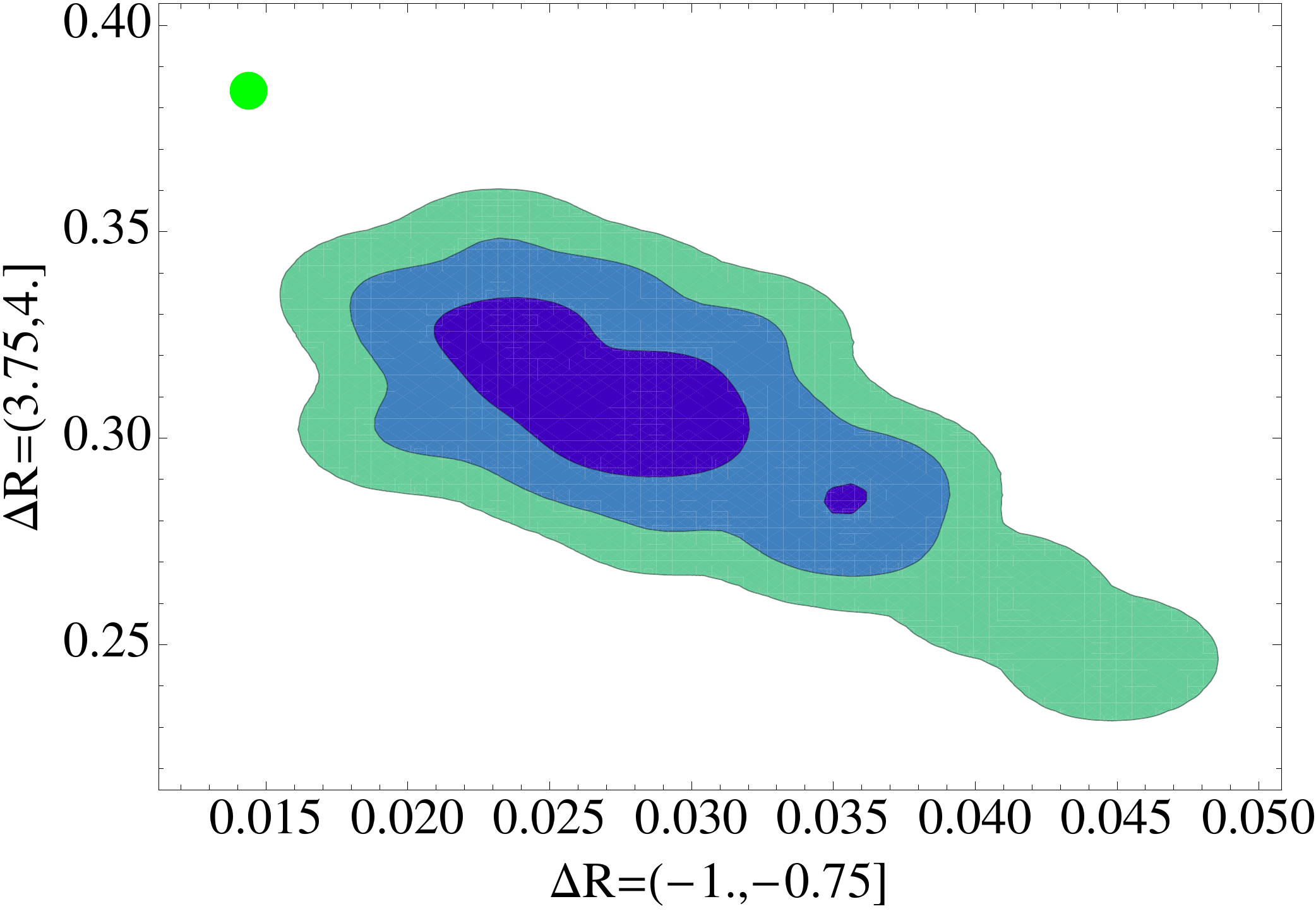}
    \caption{Same as Fig.~\ref{D1-corre} for the two robustness bins of Fig.~\ref{D2-bins}. The bin height corresponding to $\real_{\rm U+}$ is shown as a green disk. See Section \ref{Union21cuts} for more details.}
    \label{D2-corre}
\end{center}
\end{figure}
%%%%%%%%%%%%%%%%%%%%
%%%%%%%%%%%%%%%%%%%%

To further test our systematic search for systematic biases, it would be interesting to apply the internal robustness method to a dataset for which one indeed expects \emph{a priori} a significant amount of contamination due to systematics. Luckily one such sample is readily available. The Union2.1 catalogue was in fact constructed by enforcing group quality criteria to their full supernova set of 753 elements, which resulted in the removal of 173 SNIa. The criteria were~\citep{Suzuki:2011hu}:
\begin{enumerate}
    \item that the CMB-centric redshift is greater than 0.015;
    \item that there is at least one point between $-15$ and $6$ restframe
        days from B-band maximum light;
    \item that there are at least five valid data points;
    \item that the entire 68\% confidence interval for the SALT2 parameter $x_1$ lies between $-5$ and $+5$;
    \item data from at least 2 bands with rest-frame central         wavelength coverage between 2900 {\AA} and 7000 {\AA};
    \item at least one band redder than rest-frame U-band.
\end{enumerate}

Now, these 173 SNIa are precisely ones for which systematics could be a dominant factor. However, for 38 of these the lightcurve fitter algorithm did not converge, so we do not have a measurement of their distance moduli. We thus analyzed the Union2.1 catalogue augmented with 135 supernovae that did not pass their quality cuts. We show these excluded supernovae in Fig.~\ref{cuts}. We will refer to this dataset of 715 SNe as $\real_{\rm U+}$.
% and therefore did not make it into the final Union2.1 compilation

As before, we have normalized both $\real_{\rm U+}$ and relative mocks to $\bar \chi^{2}=1$. The dataset $\real_{\rm U+}$ has indeed a very high $\bar \chi^{2} = 1.72$ which, in order to go to unity, needs a quite large added error of $\sigmaextra=0.070$ magnitudes. We show the results in Fig.~\ref{D2-PDF}, which has been obtained using the Pearson distribution (for both left and right panels) as explained in Section \ref{anapdf}. As compared to the analysis of $\real_{\rm EdS}$ of Fig.~\ref{D1-PDF}, we have used a finer binning in the robustness. Now indeed the signal is in the body of the iR-PDF and not in the tail. The reason for this different behavior is that generally, by decreasing the value of $\bar \chi^{2}$, one makes the iR-PDF more peaked and with a shorter low-robustness tail. Therefore, in order to put the large $\bar \chi^{2}$ of $\real_{\rm U+}$ to unity, the latter catalogue has been ``over normalized'', probably because the high value of $\bar \chi^{2} = 1.72$ was due to just a bunch of very biased supernovae. From the results of Fig.~\ref{D2-PDF} we can claim a detection at more than $3\sigma$ of the catalogue as being systematics-driven, which is a completely independent check that the SNe that did not pass the quality cuts were indeed dominated by systematic effects.

We show in Fig.~\ref{D2-bins} the distributions of the bin heights for two robustness bins that have a strong exclusion signal.
We have plotted both the fitted Pearson distribution used in Fig.~\ref{D2-PDF} and the fitted gaussian distribution for comparison.
The $\real_{\rm U+}$ datum relative to the bin $(3.75,4]$ lies at 5.4-$\sigma$ confidence levels. By estimating the uncertainty on the exclusion signal as explained at the end of Section \ref{anapdf}, we find that the signal is $>3.8\sigma$ at 95\% confidence level.

Finally, in Fig.~\ref{D2-corre} we show the contour plot of the correlated binned distribution of bin heights for the two robustness bins of Fig.~\ref{D2-bins}.
As before, we show this figure to illustrate the correlation between bins with strong signal, but we do not use it to assess the statistical properties of $\real_{\rm U+}$ as we only have a limited sample of mock catalogues.

%%%%%%%%%%%%%%%%%%%%%%%%%%%%%%%%%%%%
\subsection{Union2.1 dataset}
\label{Union21}
%%%%%%%%%%%%%%%%%%%%%%%%%%%%%%%%%%%%

%%%%%%%%%%%%%%%%%%%%
%%%%%%%%%%%%%%%%%%%%
\begin{figure*}
\begin{center}
    \includegraphics[height= 6.15 cm]{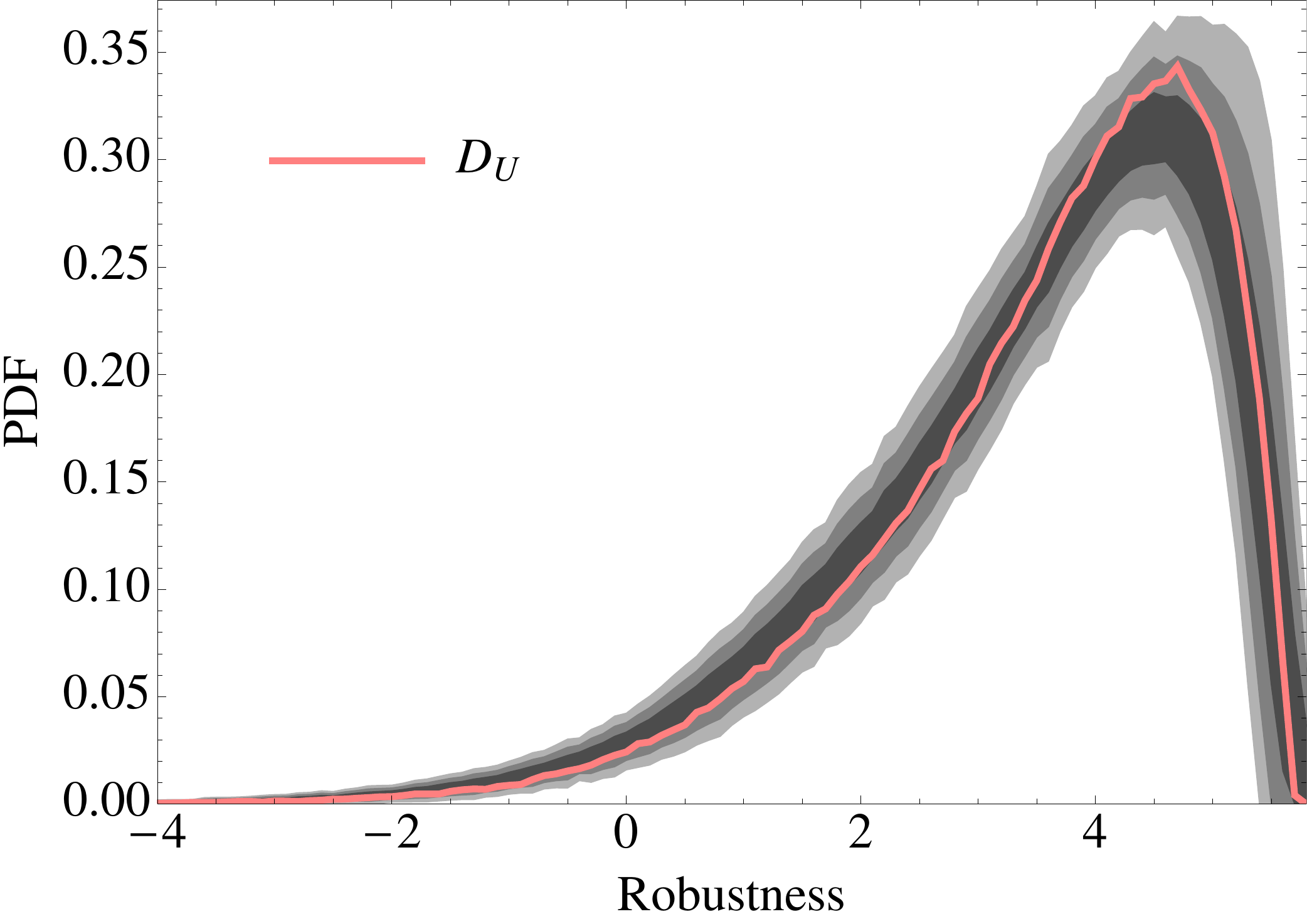}
    \quad
    \includegraphics[height= 6.15 cm]{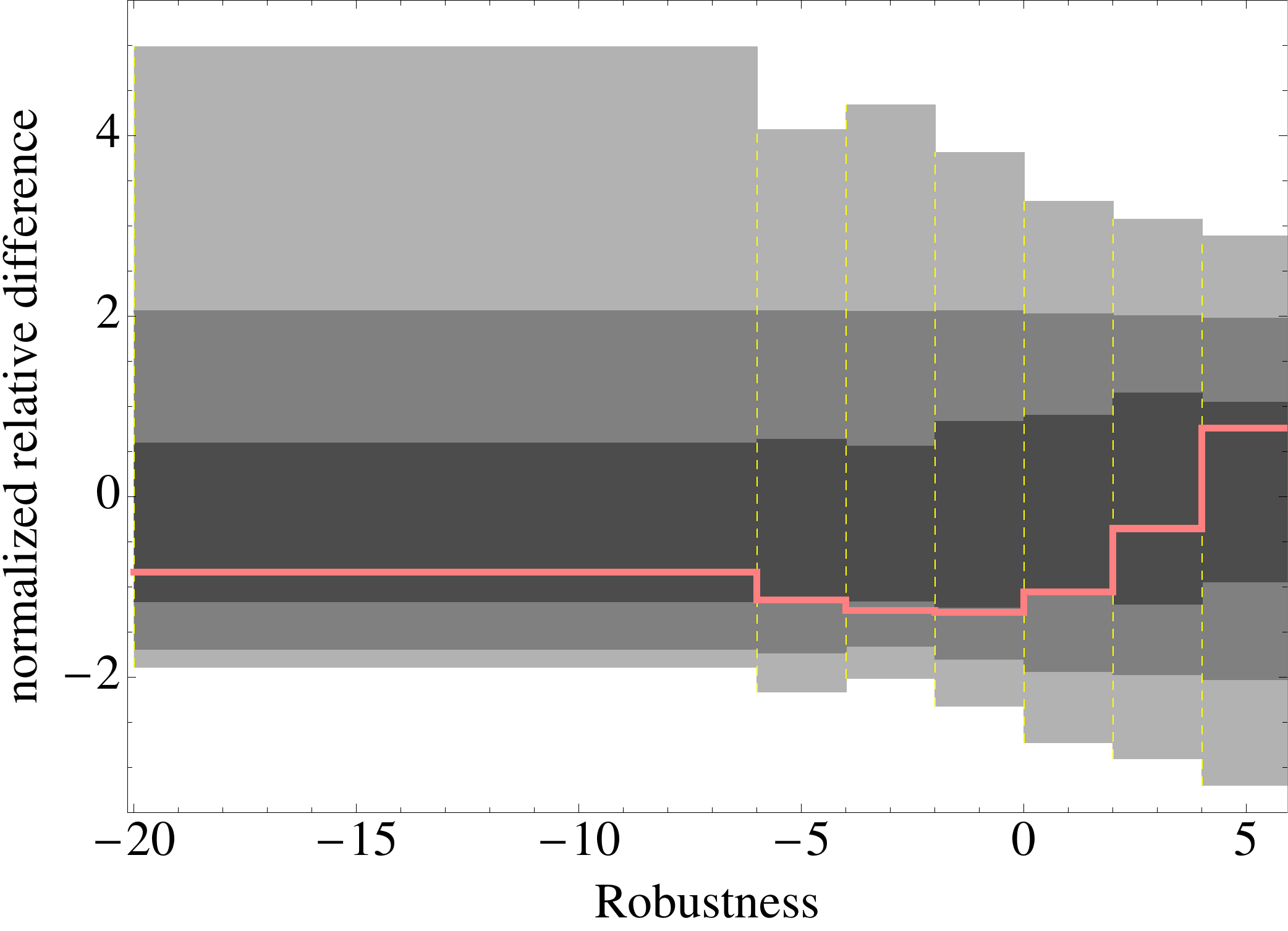}
    \caption{Same as Fig.~\ref{D1-PDF} for the dataset $\real_{\rm U}$ (pink curve). These plots show that we did not find any sign of systematic effects in the Union 2.1 catalogue.  See Section \ref{Union21} for more details.}
    \label{D3-PDF}
\end{center}
\end{figure*}
%%%%%%%%%%%%%%%%%%%%%
%%%%%%%%%%%%%%%%%%%%%

%%%%%%%%%%%%%%%%%%%%
%%%%%%%%%%%%%%%%%%%%
\begin{figure*}[t!]
\begin{center}
    \includegraphics[height= 6.3 cm]{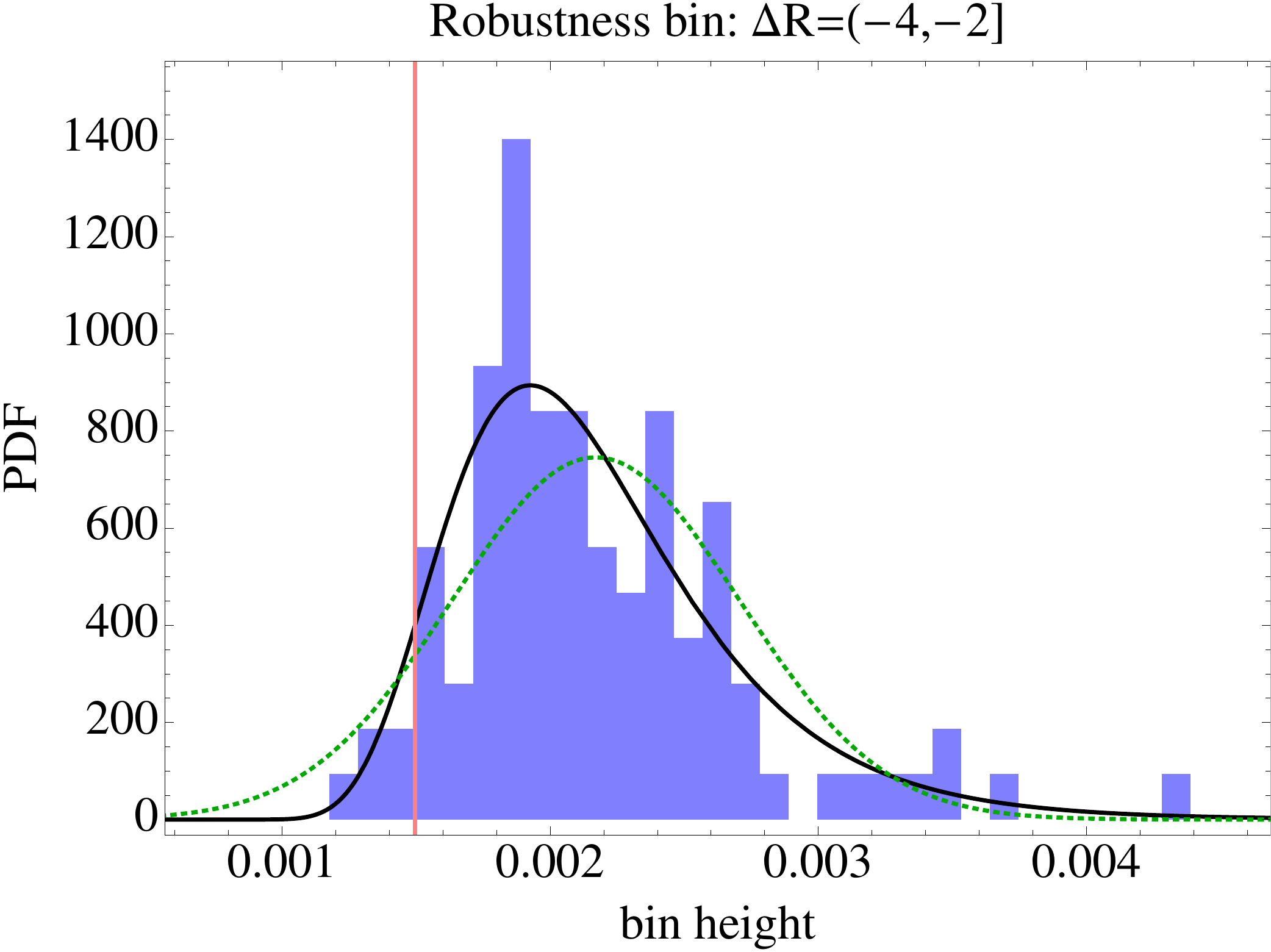}
    \qquad
    \includegraphics[height= 6.3 cm]{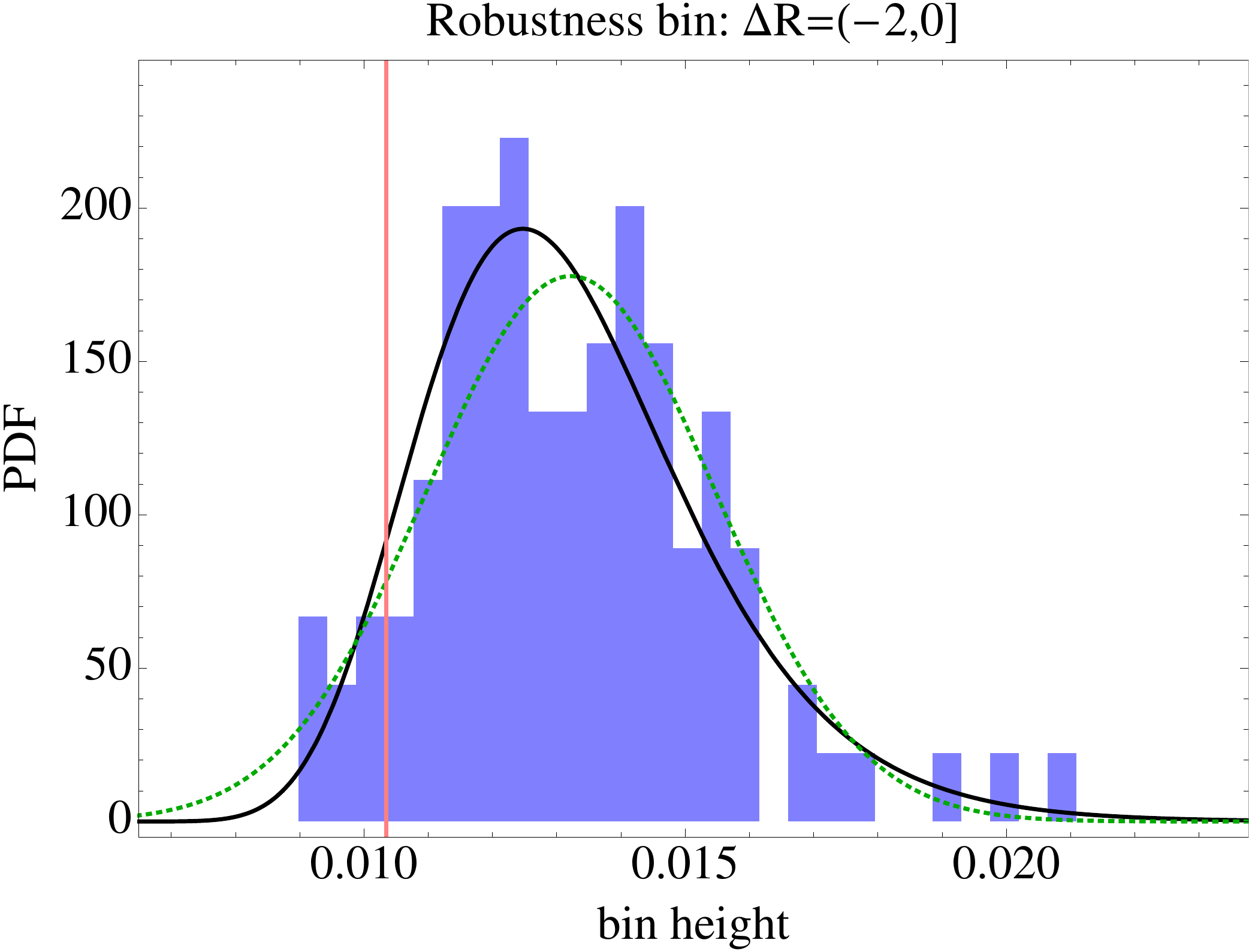}
    \caption{Same as Fig.~\ref{D1-bins} for the dataset $\real_{\rm U}$.  The bin height corresponding to $\real_{\rm U}$ is shown as a pink vertical line, and corresponds to the pink curve shown on the right panel of Fig.~\ref{D3-PDF}. These two robustness bins were chosen for this plot as they have the strongest (even though still weak) signal. See Section \ref{Union21} for more details.}
    \label{D3-bins}
\end{center}
\end{figure*}
%%%%%%%%%%%%%%%%%%%%%
%%%%%%%%%%%%%%%%%%%%%

Now that we have tested the sensitivity of the internal robustness method, we will show the results for the actual Union2.1 catalogue, to which we will refer as $\real_{\rm U}$. Fig.~\ref{D3-PDF} shows that we did not find any sign of systematic effects in the Union2.1 compilation. As Union2.1 is constructed from a collection of different instruments, it is a valid concern whether systematics are properly accounted for. For instance, individual estimations for the Malmquist bias in each of the supernova sub-samples were not made~\citep{Kowalski:2008ez} and in general all systematics were treated by adding nuisance parameters~\citep{Suzuki:2011hu}. Moreover, their data in the compact form here used, which provides directly the distance moduli, assumes a particular cosmological model (the SALT2 parameters $\alpha$ and $\beta$ are fixed in the best fits values given by such model) and is thus not technically independent data, as instead we implicitly assumed here. Our results are nevertheless clear, and serves as a cross-check on this compilation. This is one of the main results of this work.

\begin{figure}
\begin{center}
    \includegraphics[width=\columnwidth]{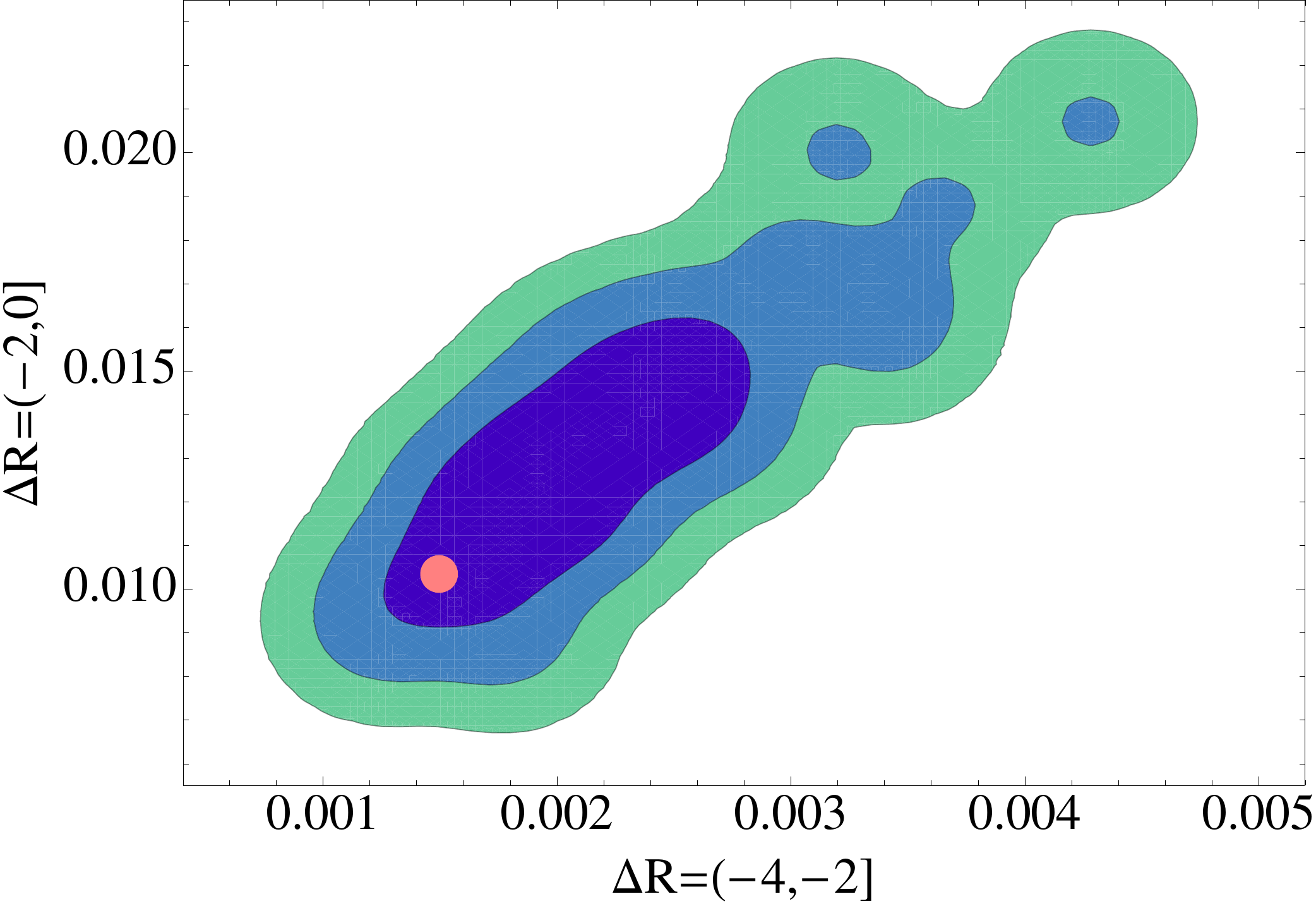}
    \caption{Same as Fig.~\ref{D1-corre} for the two robustness bins of Fig.~\ref{D3-bins}. The bin height corresponding to $\real_{\rm U}$ is shown as a pink disk. See Section \ref{Union21} for more details.}
    \label{D3-corre}
\end{center}
\end{figure}

As in the previous Sections, we show in Fig.~\ref{D3-bins} the distributions of the bin heights for two robustness bins that have the strongest signal. Again, we have plotted both the fitted Pearson distribution used in Fig.~\ref{D3-PDF} and the fitted gaussian distribution for comparison. Note that, similarly to Fig.~\ref{D1-PDF}, in the left panel of Fig.~\ref{D3-PDF} the gaussian distribution has been used. The latter is indeed a good template PDF for the robustness bins in the body with values $0<R<6$. In Fig.~\ref{D3-corre} we show the contour plot of the correlated binned distribution of bin heights for the latter two robustness bins. As before, we show this figure to illustrate the correlation between bins with strong signal.

%%%%%%%%%%%%%%%%%%%%
%%%%%%%%%%%%%%%%%%%%
%%%%%%%%%%%%%%%%%%%%
%%%%%%%%%%%%%%%%%%%%

%%%%%%%%%%%%%%%%%%%%%%%%%%%%%%%%%%%%
%%%%%%%%%%%%%%%%%%%%%%%%%%%%%%%%%%%%

\section{Conclusions}

\label{sec:conco}

%%%%%%%%%%%%%%%%%%%%%%%%%%%%%%%%%%%%
%%%%%%%%%%%%%%%%%%%%%%%%%%%%%%%%%%%%

In this paper we proposed a new Bayesian statistical method, dubbed Internal Robustness, to detect the presence of systematics with an automated procedure. While a clear physical understanding of all effects contributing to any cosmological observation is our ultimate goal, we are still far from that level and must thus resort to statistical inference to guide us. Once identified, systematic-contaminated data can be further analyzed and eventually corrected (or at least excluded). Fig.~\ref{C-EdS} nicely summarizes the ultimate goal of internal robustness, that is, to go from the original set in green to the ``decontaminated'' set in blue. The contours do not look significantly different (i.e.~same precision), but their position has substantially moved of $\sim 3 \sigma$ (i.e.~much better accuracy).

In order to test this method we applied it to three supernova catalogues, all based on the Union2.1 dataset: the artificial $\real_{\rm EdS}$ (with 100 mock supernovae drawn from the ``wrong'' EdS model); the dataset $\real_{\rm U+}$ which includes supernovae that were previously excluded (because they were suspected of being dominated by systematics); and the ``normal'' catalogue $\real_{\rm U}$ (the standard Union2.1). The latter, nevertheless, is constructed from an assortment of different instruments, both ground and space based, and it is not completely clear \emph{a priori} whether it would exhibit signs of (leftover) systematic effects.

Our results clearly show no evidence for systematic contamination in the standard Union2.1 data. This is an important result in itself and it serves as a cross-check on this very heterogeneous SNe compilation. For the other two catalogues, the results are also telling.
Using a Pearson template to fit the distribution of mock data, we find robustly that the $\real_{\rm EdS}$ and $\real_{\rm U+}$ catalogues are systematics-driven at over 3.3-$\sigma$ and 3.8-$\sigma$ confidence level, respectively.

In the present paper we have focused on assessing if a given dataset is dominated by systematics or not.
If a catalogue does not pass the internal robustness test, then the next step would be to find the subsets of data that are most probably contaminated.
To accomplish the latter one needs an efficient algorithm to find partitions with values of internal robustness as low as possible.
Once the contaminated data are found they could be used to help test some suggested hypothesis, such as the existence of two distinct supernova populations, the evolution of the supernova intrinsic luminosity with redshift and/or host metallicity \emph{et cetera}.
We will address this issue in forthcoming work.

Although we selected supernovae as our testbed, the Internal Robustness method could be applied to any data, especially if one has reasons to suspect that the data come from two or more distinct populations. The scope of application is analogous to the standard goodness-of-fit tests, which the proposed method generalizes, and we are curious to see what it will tell when applied to other observables.

\section*{Acknowledgments}

It is a pleasure to thank Juan Garcia-Bellido, Caroline Heneka, Natallia Karpenka, Richard Kessler, Bruno Lago, Michele Maltoni, Savvas Nesseris, Ribamar Reis, Ignacy Sawicki and Wessel Valkenburg for fruitful discussions.
LA and VM acknowledge support from DFG through the TRR33 program ``The Dark Universe''. MQ is grateful to Brazilian research agency CNPq for support and to ITP, Universität Heidelberg for hospitality during part of the development of this project.

%%%%%%%%%%%%%%%%%%%%%%%%%%%%%%%%%%%%
%%%%%%%%%%%%%%%%%%%%%%%%%%%%%%%%%%%%
\bibliographystyle{mn2e}
\bibliography{syst-bias}
%%%%%%%%%%%%%%%%%%%%%%%%%%%%%%%%%%%%
%%%%%%%%%%%%%%%%%%%%%%%%%%%%%%%%%%%%

\appendix

\section{Technical Details}

\label{app:details}

While scanning supernova subsets and building the robustness histogram,
one has to carefully analyze small subsets. These small
subsets produce indeed contours which are not only very broad but also have
best-fit values very far from typical ones obtained with the full catalog.
In the $\{\omegam,\omegal\}$ parameter space this entails
some technical problems.

One issue has to do with ``no Big-Bang'' scenarios.
For any positive value of $\omegam$ there exists a positive value
of $\omegal^{{\rm max}}$ for which the Hubble function
$H(z)$ goes to zero at a finite redshift. This $\omegal^{{\rm max}}(\omegam)$
line is usually referred to as the no-Big-Bang line (or ``loitering''
line), values of $\omegal$ above which are excluded. The
same issue arises for any negative value of $\omegam$, although
this is usually correctly perceived as an unimportant technicality
as a negative $\omegam$ is physically excluded \emph{a priori}.
Here, on the other hand, allowing negative $\omegam$ is reasonable
as it will be a hint that systematics are involved (we may be misinterpreting
a systematic parameter for $\omegam$ -- see Section~\ref{sec:syst-param}).
Likewise, a likelihood that embraces values of $\omegal>\omegal^{{\rm max}}$
may also be a hint of hidden systematics.

\begin{figure}
\begin{center}
    \includegraphics[width=\columnwidth]{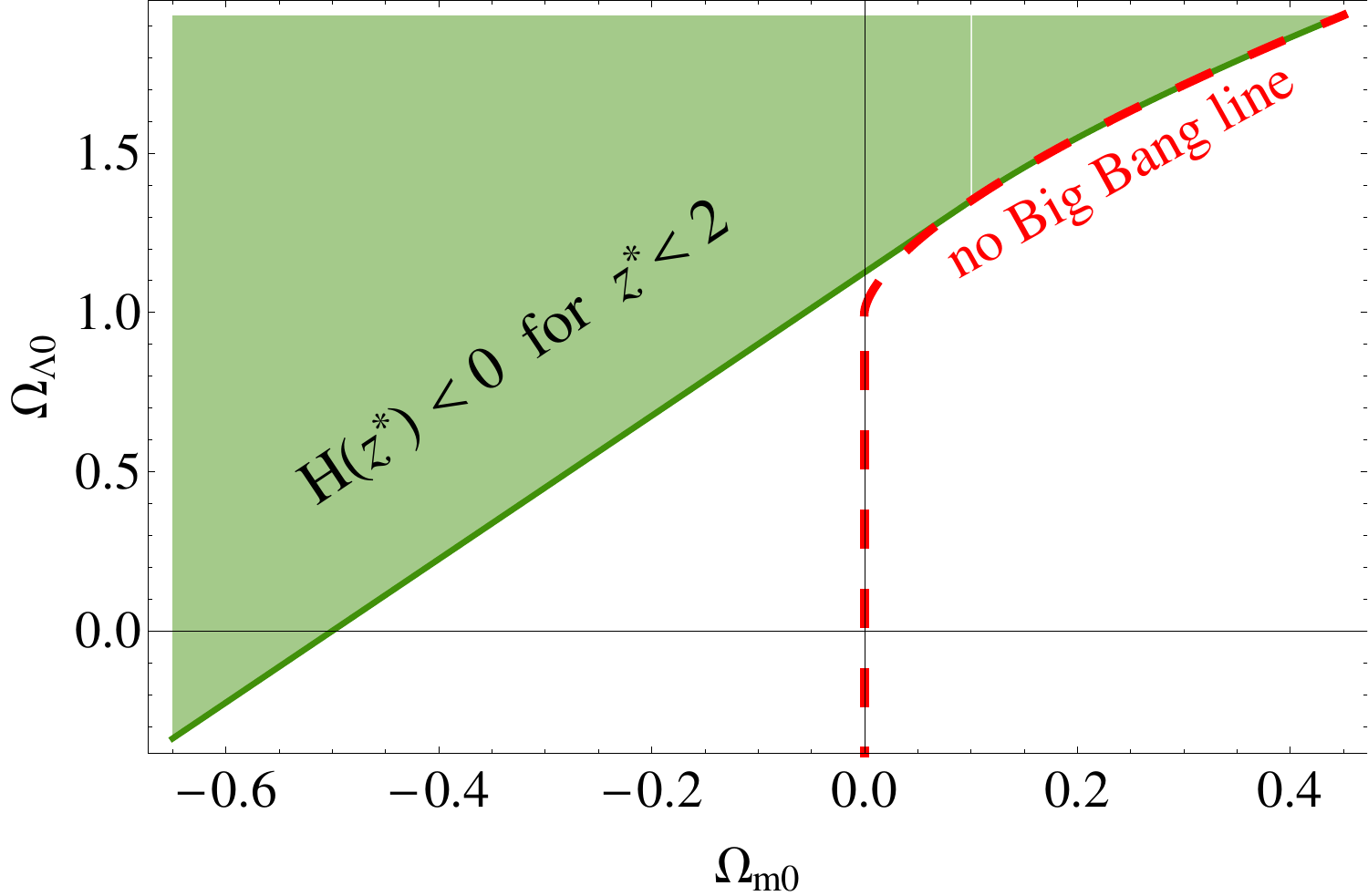}
    \caption{Shown is the excluded parameter-space region (green shaded area) for which the expansion rate is negative for $z<2$. This ensures that the distance (proportional  to $\int\rd z/H(z)$) to a source at redshift $z<2$ is not singular. This is a relaxation of the usual no-big-bang region (left of the
    red dashed line). See Appendix~\ref{app:details} for more details.}
    \label{fig:nobigbangz2}
\end{center}
\end{figure}

In order to allow as much freedom as possible to the likelihood contours,
we adopted an ``extended'' no-Big-Bang line: we allow values
of $\{\omegam,\omegal\}$ that do not exhibit a singularity
in $H(z)$ for $z<2$. This encompasses all supernovae measured so
far and opens up a large region for which $\omegam<0$; see Fig.~\ref{fig:nobigbangz2}.

To find the line that delimits such a region one computes first the
solutions $\omegal^{{\rm max}}(z,\omegam)$ for which
$H(z)=0$. The minimum of $\omegal^{{\rm max}}(z,\omegam)$
with respect to $z$ gives the lower redshift for which $H(z)=0$
for a given $\omegam$. The function we are looking for is then
$\omegal^{{\rm max}}\big(z^{\star}(\omegam),\omegam\big)$
where $z^{\star}(\omegam)$ is either the redshift corresponding
to the minimum $z_{{\rm m}}(\omegam)$ or 2, whichever is lower.
In other words, $z^{\star}={\rm Min}(z_{{\rm m}},2)$. Carrying out
this calculation one gets
\begin{align}
\omegal^{{\rm max}}(\omegam)=\left\{ \begin{array}{cc}
\frac{9}{8}(1+2\omegam), & \quad\omegam\le1/10\,;\\
\omegal^{{\rm noBB}}(\omegam), & \quad\omegam\ge1/10\,;
\end{array}\right.\label{eq:nobigbang-z2}
\end{align}
 where $\omegal^{{\rm noBB}}$ is the traditional no Big
Bang function:
\begin{align*}
\omegal^{{\rm noBB}}(x)\!= & \Big[4x^{4}+4x^{2}(1+y)-4x^{11/3}(1-x+y)^{1/3}\\
 & \;+x^{4/3}y(1-x+y)^{2/3}+4x^{10/3}(1-x+y)^{2/3}\\
 & \;-4x^{3}(2+y)-x^{7/3}(2+y)(1-x+y)^{2/3}\\
 & \;+x^{8/3}(1-x+y)^{1/3}(4+3y)\\
 & \;+\left(x^{4}\left(x^{2}+2(1+y)-2x(2+y)\right)\right)^{1/3}\Big]\\
 & \!\!\!\!\!\!\times\Big[x^{4/3}(1-x+y)^{1/3}\Big(x^{4/3}+(x(1-x+y))^{2/3}\\
 & \qquad+\left(x^{2}+2(1+y)-2x(2+y)\right)^{2/3}\Big)\Big]^{-1}\,,
\end{align*}
 where in turn $\, x=\omegam\,$ and $\, y\equiv\sqrt{1-2x}$.

\bsp

\label{lastpage}

%%%%%%%%%%%%%%%%%%%%%%%%%%%%%%%%%%%%
%%%%%%%%%%%%%%%%%%%%%%%%%%%%%%%%%%%%
\end{document}